# Numerical Analysis of Lensless Imaging with Active Metasurfaces and Single-Pixel Detectors

Julie Belleville, Prachi Thureja, and Harry A. Atwater*

Thomas J. Watson Laboratories of Applied Physics

California Institute of Technology

Pasadena, CA 91125

jbellevi@caltech.edu, pthureja@caltech.edu, haa@caltech.edu

*Contact information for corresponding author: haa@caltech.edu

**ABSTRACT**

We introduce a conceptual framework for a lensless imaging system which employs an active metasurface as a high-frequency, continuously tunable amplitude and phase modulation aperture, coupled to a discrete single-pixel detector. Using an array factor formalism, we first study fundamental limits in information collection, offering a comparison to existing technologies. We also study the effects of modulation rate and losses on the system acquisition time and signal-to-noise ratio, which place bounds on system performance for set illumination conditions. Considering both an ideal metasurface and the phase and amplitude modulation characteristics of an experimentally realized indium tin oxide-based metasurface operating at 1510 nm, we then simulate image recovery with ~60,000 image points for a 0.2 mm x 0.2 mm active metasurface aperture. We show that aberrations appearing in the simulated images produced by the metasurface can be corrected through post-processing. We further investigate trade-offs between image acquisition time and image quality both through the realization of Hadamard coupling bases and by modifying the k-space width coupling to the detector. Finally, we discuss the technical challenges which remain to be overcome for experimental realization of a lensless single-pixel imaging technology.

**INTRODUCTION**

The performance of conventional imaging systems is dictated by tradeoffs between device size, resolution, and field of view (FOV). For example, in a pinhole camera using a fixed sensor array aperture size and a fixed distance between the object and the image detector, the captured image resolution can only be increased by increasing the distance between the pinhole and detector, which in turn decreases the field of view (FOV). The diffraction-limited resolution of conventional lens-coupled detector arrays (Fig. 1a) can be improved by increasing the numerical aperture, but this improvement is often linked to increased system complexity or aberrations at wide FOV[1]. In systems such as these which passively collect light, the sensor array imposes a critical limit on the number of resolvable points. At the Nyquist limit, a passive system with N pixels along a sensor axis can resolve at most N/2 points along said axis. For a given sensor array, this leads to a fundamental tradeoff between resolution and FOV, regardless of how light is acquired.

Further, lens-coupled image sensors require a minimum optical 'thickness', related to the geometrical configuration of the information channels in the system[2]. This optical thickness requirement in turn directly dictates how present-day 'thick' optical systems are designed and manufactured, and for which optical component assembly costs often dominate the total optical system cost. This has led to an increased interest in lensless imagers (Fig. 1b) in which an optical encoder is placed in close proximity to a sensor to enable computational reconstruction[3,4].

The concept of single-pixel (single-detector) imaging has emerged in the last decade to address the dependency of conventional imaging on large sensor arrays[5]. While visible light imaging has access to megapixel-array CMOS sensors, imaging systems at other wavelengths typically suffer from more limited resolution and higher cost and complexity due to immature detector technology. The cost, size, and operating temperatures of detectors hinders applications such as cancer diagnosis, artwork inspection, and semiconductor wafer inspection[6]. Single-pixel imaging uses an actively modulated aperture to collect the light scattered by the scene and incident on a detector. This allows the system to utilize the often superior performance (SNR, dynamic range, bandwidth) of single-pixel detectors while preserving spatial resolution, at the cost of longer acquisition times[7]. An exemplary configuration is illustrated in Fig. 1c. A digital micromirror device (DMD) is placed in the image plane of a lens and acts as a binary (on/off) amplitude modulator which directs light at each micromirror either towards or away from a single detector. By displaying orthogonal basis elements $O_t$ across the DMD at different times $t$, an image can be recovered by summing the elements $O_t$ weighted by the measurement at time $t$. In some cases, a compressed sensing basis may replace the orthogonal basis, significantly reducing the number of measurements required to reconstruct a scene[8].

We develop here an approach to imaging using active metasurfaces as modulators in lensless single-pixel detector-based systems (Fig. 1d). Metasurfaces are planar arrays of subwavelength scatterers which allow unprecedented and compact control of degrees of freedom in light including amplitude, phase, polarization[9]. Passive metasurfaces have emerged as a promising platform to improve imaging systems as lightweight alternatives to refractive lenses[10,11], as means to achieve planar, lensless wide FOV imaging[12], and as polarization cameras[13]. Whereas passive metasurface scatterers have fixed optical properties, the refractive index of active metasurface scatterers can be dynamically tuned by external stimuli, allowing reconfigurable control over the wavefronts of light. Scatterers can be tuned through a variety of methods such as mechanical actuation[14], liquid crystal reorientation[15], carrier accumulation[16], and the electro-optic effect[17]—each offering different tradeoffs in terms of technological maturity, operation wavelength, operation mode (reflection or transmission), and modulation speeds[9]. With proper choice of material platform, active metasurfaces can achieve highly desirable characteristics for imaging such as high modulation speeds (>10 MHz) to reduce image acquisition time, compact sizes, wide

FOV, and continuous-valued phase and amplitude modulation[18]. In contrast, DMDs are limited to binary amplitude modulation at ~30 kHz and must be used jointly with appropriate lenses to achieve wide FOVs[19].

The active metasurface-based single-pixel imaging device architecture is conceptually described in Fig. 1d. The active metasurface element selectively couples light from the far field into a single detector element, replacing both the lens and modulator elements most commonly present in single-pixel imaging setups. The coupling between different incident wave vectors and the detector is controlled through the configuration of each scatterer (metasurface configuration) and can be reconfigured at a modulation rate dictated by the modulation bandwidth of the metasurface, allowing spatial information to be resolved. We focus on electrical tuning mechanisms, because of their potential for subwavelength, independent addressing of scatterers[9], and so the metasurface configuration can also be defined as the set of applied voltages. Here, the detector is schematically illustrated as either in the transmissive or reflective far field of the metasurface. While reflective active metasurfaces are more developed at this time, active transmissive architectures[20] are preferable for minimizing device footprints as we anticipate that for many detector technologies, near field coupling between the active metasurface and detector would be possible in that configuration[12]. An active metasurface coupled to a single-pixel detector in the near field has 'zero' added thickness, and so the system thickness is defined by the cumulative thickness of the active metasurface and the detection. This could lead to compact, low-profile single-pixel detector systems with extremely low size, weight and power consumption (SWaP). Whereas in lens-coupled imagers the ratio of lens diameter to focal length limits the FOV, the FOV of an active metasurface imager is limited only by scatterer pitch and the angular sensitivity of scatterer amplitude and phase responses.

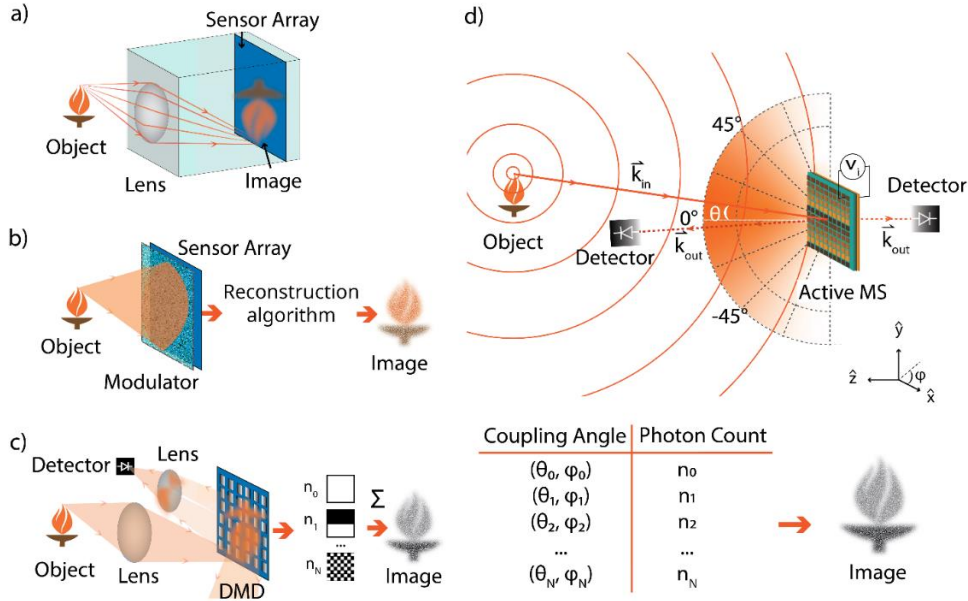

**Figure 1. Imaging systems**. (a) A standard lens-coupled sensor array imaging system. Light is focused onto the image plane through a lens, and collected at each point by a sensor array. (b) A lensless imaging system. Information from the scene is encoded onto a sensor array through a modulator placed directly against the array. The information collected by the sensor array must be computationally decoded to produce an image. (c) Single-pixel imaging system, using a digital micromirror device (DMD) as an amplitude modulator. The DMD is placed at the image plane and selectively directs light from each pixel either toward or away from a single detector. (d) Proposed active metasurface single-pixel lensless imaging device. Light is incident from the far field onto an active metasurface, which selectively couples inbound light as a function of wavevectors $\vec{k}_{in}$ to a single-pixel detector, schematically represented at the outbound wavevector $\vec{k}_{out}$. The coupling can be dynamically modified via applied stimulus (depicted here as a voltage) on each metasurface scatterer.

We begin by introducing our formalism and setting fundamental limitations on the resolution, field of view (FOV), acquisition times, and signal-to-noise ratio (SNR) of the proposed system. Next, we simulate imaging through different measurement bases and discuss effects of non-ideal behaviour on the point-spread function of the system. We also consider how the k-space width of detector coupling can improve SNR at the cost of resolution. Finally, we provide an outlook on the challenges and path towards commercially useful realization of such imaging systems.

**Single pixel active metasurface imaging; simulation formalism and assumptions**

In this work, we model the imaging capabilities of a single detector coupled to a local, two-dimensional active metasurface composed of scatterers placed in a rectangular grid. We consider the near-monochromatic imaging of a far-field scene and assume no significant interference between different points of a scene over the integration time of our detector. For simplicity, we assume a single polarization. Under these assumptions, the photon flux per unit k-space arriving at a detector, $I_{detected}(\vec{k}_{out})$, coupled to the outbound wavevector $\vec{k}_{out} = k_{out,x}\hat{x} + k_{out,y}\hat{y}$ (taken to be $\vec{k}_{out} = \vec{0}$ in our simulations) can be written as

$$I_{detected}(\vec{k}_{out}) = \int_{\frac{\vec{k}_{in}}{|\vec{k}_{in}|}\leq 1} I_{in}(\vec{k}_{in})|G(\vec{k}_{out},\vec{k}_{in})|^2 |A(\vec{k}_{in}-\vec{k}_{out})|^2 d\vec{k}_{in} \quad (1)$$

where $I_{in}(\vec{k}_{in})$ is the photon flux per unit k-space arriving at the metasurface, $G(\vec{k}_{out},\vec{k}_{in})$ describes the complex field coupling between inbound wavevector $\vec{k}_{in}$ and outbound wavevector $\vec{k}_{out}$ (see Fig. 1d) for a single metasurface scatterer, and $A(\vec{k}_{in}-\vec{k}_{out})$ is the array factor

$$A(\vec{k}_{in}-\vec{k}_{out}) = \sum_n a_n e^{i\psi_n} e^{i(\vec{k}_{in}-\vec{k}_{out})\cdot\vec{r}_n} \quad (2)$$

where $\vec{r}_n = (x\hat{x} + y\hat{y})$ is the position of scatterer $n$, and $a_n(v_n)$ and $\psi_n(v_n)$ are the amplitude and phase response, respectively, subject to voltage $v_n$. This derivation is provided in Supplementary Information Part 1 (SI.1). The array factor is the Fourier transform of the metasurface configuration and encodes the far field interference between uncoupled scatterers. We assume that the phase response and material losses of scatterers can be approximated as angle-independent over the FOV of interest. This assumption is generally valid across a wide FOV for metasurfaces with a lower quality factor, as in the case of an exemplary plasmonic TCO metasurface discussed later in this work. The coupling from inbound to outbound wavevector is $G(\vec{k}_{out},\vec{k}_{in}) = g(\vec{k}_{out})g(\vec{k}_{in})$, the product of coupling from an inbound wavevector into the scatterer mode(s) and the reciprocal coupling from the scatterer to an outbound wavevector. The coupling coefficient between scatterer and plane waves is the antenna factor, $g$. In this work, we adopt the convention of indexing our plane waves by their in-plane wavevector components only, because the normal wavevector component $k_z$ is fully constrained by our near-monochromatic assumption. While our formalism is also suitable for the analysis of multi-pixel systems, we leave this analysis to future work.

To simplify the discussion that follows, we assume that our detector receives light from a narrow acceptance angle range around a nominal angle $\theta = 0°$ (Fig. 1d) for an integration time $\tau$, such that the number of detected photons $N_{detected}$ is

$$N_{detected} = \tau\, Q_e (\Delta k_D)^2 L^2 \int_{\frac{\vec{k}_{in}}{|\vec{k}_{in}|}\leq 1} I(\vec{k}_{in})|G(\vec{0},\vec{k}_{in})|^2 |A(\vec{k}_{in})|^2 d\vec{k}_{in} \quad (3)$$

where $(\Delta k_D)^2$ indicates the k-space area (acquisition angle) of the detector and is constant across all measurements and $Q_e$ is the quantum efficiency, the ratio of incident photons to collected charge carriers. The quantum efficiency more generally can be replaced by any measure of detector efficiency.

**RESULTS**

**Fundamental bounds on information content**

The simplest method of imaging a scene using an active metasurface is to retrieve the scene point-by-point. The role of the active metasurface in this system is then to direct light from time-dependent target solid angles toward the normal. This task is notably analogous and reciprocal to the well-studied problem of beam-steering[21], where incident light is directed to another angle. Conceptually, this approach corresponds to imaging of the scene with a basis of delta functions each aimed at collecting light from a target point in k-space $\vec{k}_t$. At each time, we select a target point $\vec{k}_t$ and set the metasurface configuration such that the square of the array factor can be approximated as $\left|A(\vec{k}_{in})\right|^2 = |A_{\vec{k}_t}|^2 \delta(\vec{k}_{in} - \vec{k}_t)$. The quantity $|A_{\vec{k}_t}|^2$ is the proportion of coupled power which scatters back into plane waves for measurement of $\vec{k}_t$ and accounts for absorption losses in the metasurface. Under this assumption, Eqn. 3 simplifies further to

$$N_{detected}[\vec{k}_t] = \tau\, Q_e (\Delta k_D)^2 L^2\, \eta_{\vec{k}_t}\, I_{in}(\vec{k}_t) \tag{4}$$

The square brackets denote the fact that $N_{detected}$ is not a continuous function of $\vec{k}_t$ but rather the number of photons obtained by a discrete measurement. We define the metasurface scattering efficiency $\eta_{\vec{k}_t} = |A_{\vec{k}_t}|^2 \left|G(\vec{0}, \vec{k}_t)\right|^2$, which is the ratio of incident power from $\vec{k}_t$ to scattered power and accounts for both losses due to light failing to couple into the metasurface and absorption losses.

As desired, we find that the number of detected photons is directly proportional to the incident photon flux from the target point. To properly reconstruct an image, however, we must characterize the efficiency at each measurement and normalize the collected value. This can be experimentally done by taking measurements under known illumination. We call the normalized measurement $N_{ideal}$, formulated as

$$N_{ideal}[\vec{k}_t] = \left(\eta_{\vec{k}_t}\right)^{-1} N_{detected}[\vec{k}_t] \tag{5}$$

In practice, our metasurface cannot generate an ideal delta function of coupling. We therefore simulate the detected power using Eqn. 2 and 3. Each measurement yields the inner product between the scene and the intensity coupling, where the intensity coupling is set by the choice of metasurface configuration. While the reported simulations assume fully 2D addressable metasurfaces, imaging schemes such as point-by-point imaging, where the desired coupling can be written as the convolution of two linearly independent responses in x and y, are also compatible with high-speed perimeter-control architectures[22]. In such an architecture, control signals are applied only at each row and column, resulting $2N$ rather than $N^2$ control signals for an $N \times N$ scatterer metasurface, significantly simplifying the fabrication of a large-scale aperture.

The properties of active metasurface scatterers (the dependency of scattered amplitude and phase on voltage, and the coupling coefficients to and from plane waves) generally depend on the metasurface design. To consider the fundamental resolution limit achievable by local active metasurfaces, we first consider a fully "ideal" metasurface with uniform coupling between any plane wave and the scatterer modes (isotropic antenna factor), no material loss, and $2\pi$ phase control. Deviations from this ideal are considered later in the text. By setting constant phase gradients across the metasurface (SI.2), we retrieve a coupling full width at half maximum (FWHM) in normalized k-space of

$$FWHM_i = \frac{2H\lambda}{N_i \Delta_i}, \qquad sinc(H) = \frac{1}{\sqrt{2}}, H \approx 0.443, i \in \{x, y\} \tag{6}$$

where the FWHM is given in units of $\vec{k}/|k|$, $FWHM_i, N_i, \Delta_i$ are the FWHM, scatterer number, and scatterer period or pitch along the axis $i$. This corresponds to a diffraction-limited resolution.

The geometry of the active metasurface scatterers additionally dictate the maximum FOV which can be imaged by the system. Assuming once again a subwavelength rectangular lattice of pitch $\Delta x < \lambda, \Delta y < \lambda$, we can determine the regions of unit k-space in which imaging is possible without aliasing as a function of the ratios $\frac{\lambda}{\Delta x}, \frac{\lambda}{\Delta y}$ [23]. Achievable FOVs are reported in SI.3. In the results that follow, $\frac{\lambda}{\Delta x}, \frac{\lambda}{\Delta y} \geq 2$ and a full 180° FOV is available without aliasing[23].

By considering both the k-space resolution and the FOV available to the system, we can resolve the total number of distinct, diffraction-limited points recoverable as (SI.2)

$$N_p = \frac{\pi \cdot NA^2}{4H^2} \left(\frac{N_x \Delta_x}{\lambda}\right) \left(\frac{N_y \Delta_y}{\lambda}\right) \tag{7}$$

Throughout this paper, we will consider an example active metasurface operating at a wavelength $\lambda = 1510 nm$ with a pitch $\Delta_x = \Delta_y = 400\ nm$. This choice is motivated by availability of data for an experimentally demonstrated metasurface at this wavelength. For this example, assuming a $0.2\ mm \times 0.2\ mm$ sized aperture, $N_p \approx 73,400$. In contrast, a DMD of the same aperture size with a pitch of $5.4\ \mu m$[24] would have only ~1400 pixels, a short-wavelength infrared sensor array with pixels having a pitch of $5\ \mu m$[25] would have ~1700 pixels, and even a CMOS sensor array with a small pixel pitch of $1.2\ \mu m$ would have only ~29 000 distinct pixels. Thus, we see that for a set device aperture size, owing to its dense pixel array, a wide-FOV active metasurface imaging device could resolve more points than lens-coupled or DMD-coupled sensors.

**Acquisition times and signal-to-noise ratio**

Due to the serial data collection in single-pixel imaging systems, it is very important to characterize their image acquisition time. In general, the total acquisition time for an image can be limited either by the achievable modulation rate of the metasurface or by the exposure time required to achieve a target image SNR, which is affected by loss. To focus specifically on the noise introduced by the active metasurface element, we assume a shot noise limited system and define the image SNR according to Eqn. 8 as

$$SNR = \frac{\sum_{k_x, k_y} N_{ideal}[k_x, k_y]^2}{\sum_{k_x, k_y} \eta^{-1}[k_x, k_y] N_{ideal}[k_x, k_y]} \tag{8}$$

as derived in SI.4. Here, $\eta[k_x, k_y]$ and $N_{ideal}$ are the metasurface scattering efficiency and normalized photon count as discussed in Eqn. 5.

We study the regimes in which loss or modulation rates limit the acquisition time in Fig. 2a. Here, we define losses to include only coupling and absorption losses and assume a unity quantum efficiency. Critically, light which couples to the metasurface and scatters away from the detector is not considered a loss, since selectively guiding light from non-target angles away from the detector is fundamentally necessary to our imaging system. Our listed efficiencies do not, then, express the total proportion of incident light which is translated to a signal, but rather the proportion of light which could be detected.

Figure 2a assumes an average photon flux of $7.6 \times 10^{21}$ $photons/m^2/s$ (equivalent to 1 sun or $1000\ W/m^2$ of power at a wavelength of $1510\ nm$) and considers the effects of active metasurface characteristics on the SNR for a fixed collection time. It assumes a detector bin of size $FWHM_{min}^2$, a 0.2 mm x 0.2 mm metasurface operating at $\lambda = 1510 nm$, and a 57 609 recovered image points, with a point-by-point acquisition method. We first consider how losses which are uniform across all measurements affect the SNR, through the blue shaded region which indicates SNR as a function of time for systems with an efficiency between 100% and 1%. As expected, we see 10 dB of loss for each order of magnitude drop in metasurface intensity scattering efficiency. Next, we consider the case where loss varies across measurements through the dotted green lines. Losses for each measurement are sampled from a normal distribution with mean efficiency 0.5 and a standard deviation of 0 (top line) and 0.18 (bottom line), then clipped to fall within a physical range $\eta \in (\epsilon, 1-\epsilon)$, $\epsilon \ll 1$. We find that in systems where loss varies with the measurements, the SNR is dominated by the measurements with the highest loss. Despite a mean efficiency far greater than 1% across all measurements, large variations in efficiency result in a very poor SNR.

Finally, we consider an experimentally "realizable" plasmonic active metasurface on a transparent conducting oxide (TCO) platform, with covarying amplitude and phase (shown in SI.5), and 272° of phase control, capable of >10 MHz modulation rates[16,18]. This realizable metasurface was experimentally demonstrated for a one-dimensional array in reflection; here, we extrapolate the device characteristics to a two-dimensional device. State-of-the-art for two-dimensional addressing of scatterers and associated challenges are outlined in SI.6.

We additionally assume a dipole coupling (antenna factor) between plane waves and the scatterers, such that the amplitude coupling between a plane wave and the scatterers is $g(k_x, k_y) = g(\vec{k}) \propto \sqrt{1-(k_y/|k|)^2}$. This antenna factor is not obtained experimentally, but rather is selected to highlight image and SNR features which arise from non-isotropic coupling. Through these assumptions, we calculate a mean measurement efficiency $\mu_\eta = 4.2\%$ and a standard deviation of $\sigma_\eta = 0.85\%$ across a 180° FOV, which account for both coupling and absorption. With these assumed losses, the expected SNR (shown as red triangles in Fig. 2a) is ~16 dB below that of the ideally achievable SNR. Coupling losses are most significant at wide FOVs; when we consider a smaller FOV, $\mu_\eta$ increases and $\sigma_\eta$ decreases. The portions of the scene near the normal therefore have a greater SNR than those near the edges.

Given an expected photon flux, target aperture size, SNR, and image resolution, these calculations can help select which state-of-the-art active metasurface platform is most suited to a desired application. Figure 2a shows the modulation-rate-limited image collection time for different active metasurface platforms as vertical lines. The values for currently achievable modulation rates are taken from Ref. 9 for metasurface platforms and Ref. 5 for DMDs. Importantly, these values reflect modulation rates in current experimental demonstrations and primarily consider material limitations. The modulation rate achievable across a full aperture may also be affected by the scaling of the control architectures (SI.6). We find that under the illumination conditions plotted, the realizable TCO metasurface can collect an image with an SNR of less than 35 dB faster than a liquid crystal metasurface due to its faster modulation rate. However, when aiming to retrieve high SNR images (e.g., SNR > 40 dB), or assuming a lower irradiance, a liquid crystal metasurface becomes preferable due to its lower losses.

These image acquisition rates also limit the system's ability to image dynamically. To acquire video-rate imaging of 24 frames per second at the resolution assumed in Fig. 2, each image would have to be acquired in ~42 ms, a threshold shown in Fig. 2a as a dashed vertical blue line. At the chosen illumination, this corresponds to an SNR of 37 dB for an ideal active metasurface and of 21 dB for our realizable plasmonic metasurface—for this image resolution, the required modulation rates preclude the use of liquid crystal or DMD modulation platforms. Images with 5-25 dB of noise (modelled as normal noise) can be viewed in Fig. 2b, for reference.

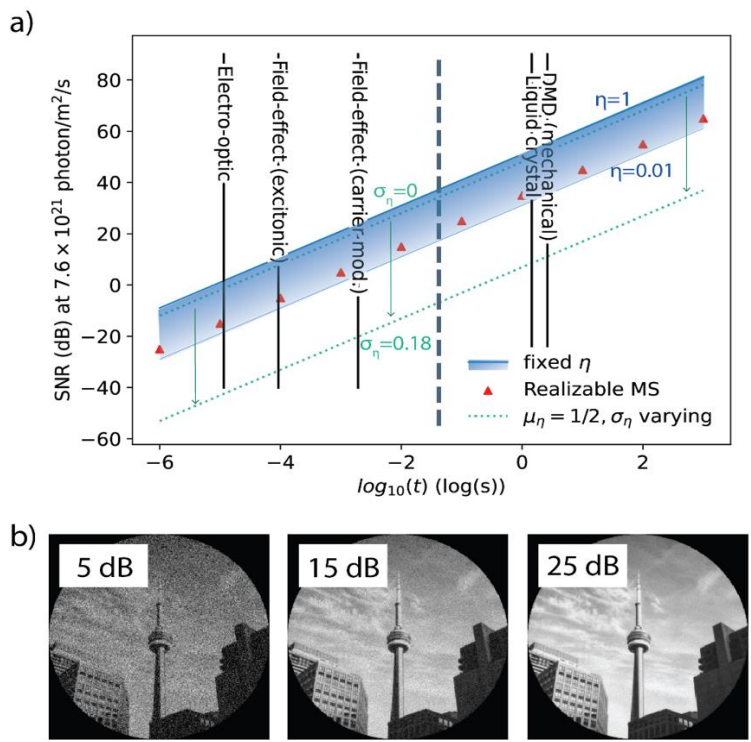

**Figure 2.** Effects of loss, SNR, and modulation rates on image acquisition time. Results assume a quasi-monochromatic light source with a photon flux of $7.6 \times 10^{21} \frac{photons}{m^2 \, s}$, an aperture size of $0.2 \, mm \times 0.2 \, mm$, and a single detector with a uniform, diffraction-limited square acceptance in k-space, and 57 609 measurements. (a) SNR (dB) as a function of the logarithm of time *t*, for various measurement loss conditions. The blue shaded region assumes a fixed efficiency across all measurements, from 1 to 0.01. The dotted green lines demonstrate the effect of variance in measurement efficiency, where the SNR at large variance is dominated by the high-loss measurements. The red triangles indicate SNR for the realizable metasurface[16]. The black vertical lines indicate the modulation-speed limited imaging time for different modulation mechanisms. The vertical dashed blue line shows the acquisition time below which video-rate imaging is achievable. (b) Example of the retrieved images at various shot-noise limited SNRs. The Poisson shot noise is approximated as normal noise.

In most experimentally realized active metasurfaces to date, it has been difficult to achieve both high modulation rates and low losses in the same device. We elected to model this specific realizable metasurface for further studies not because it is the most suitable platform for an active metasurface imaging device—platforms will lower loss would generally be preferred—but rather because we have a thorough understanding of its performance from prior work.

**Imaging of scenes with different imaging bases**

Next, we consider simulated point-by-point image collection and visually assess the image collected by active metasurfaces of modest size ($135 \, \lambda \times 135 \, \lambda$) coupled to a single detector.

We compare the performance of the previously described 'ideal' and 'realizable' metasurface models to demonstrate how strong covariation of amplitude and phase generates systematic errors in imaging. In both cases, we demonstrate diffraction-limited resolution across a FOV of nearly 180°, with neighboring image points spaced at intervals of $\frac{\lambda}{L} = 1.12\ FWHM_{min}$ apart. Here, $L$ is the metasurface dimension along one axis, $FWHM_{min}$ is the diffraction-limited FWHM. To demonstrate the feasibility of this wide FOV, we include full-wave simulations of a TCO metasurface effectively scattering light incident from up to 71° towards the normal in SI.7.

As expected, the ideal metasurface image (Fig. 3b) shows excellent agreement with the ground truth (Fig. 3a). In contrast, the realistic features assumed for our scatterers (losses, covariation of amplitude and phase, non-isotropic coupling) lead to systematic aberrations in the image (Fig. 3c). These aberrations include a decrease in the contrast of the image, as well as 'ghost images' at reflected locations which can be most clearly seen as the outline of buildings in the sky. This systematic error limits our SNR (Eqn. S27) to 15 dB, even in the absence of noise, and is studied later in the text through the point spread function (PSF) of the device.

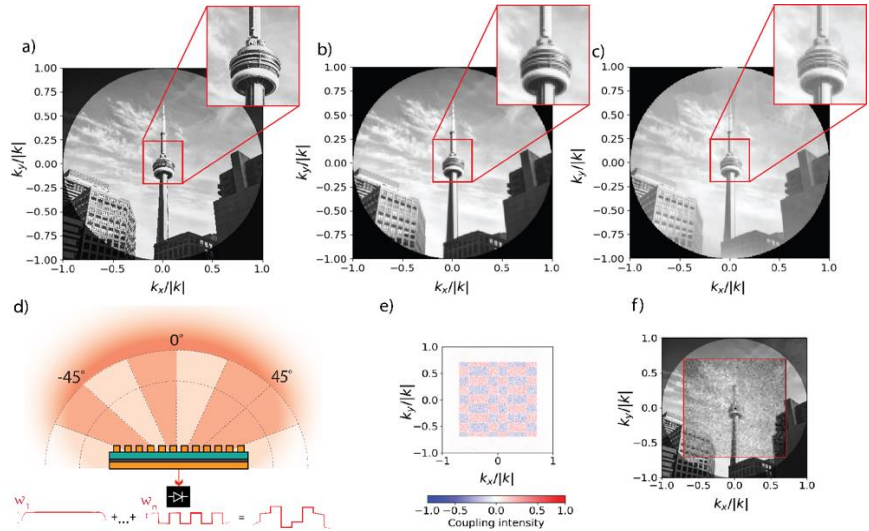

**Figure 3.** Simulated single-pixel image capture via an active metasurface imaging device.

(a) Quasi-monochromatic (approximated as a single wavelength) ground truth scene of the CN tower, defined across a full 180° field of view. Picture by Patrick Tomasso on Unsplash. (b) Simulated recovered image from an ideal active metasurface with a $511 \times 511$ scatterers, $0.2\ mm \times 0.2\ mm$ ($135\ \lambda \times 135\ \lambda$) aperture. 57 609 separate angular measurements are resolved in series, with diffraction-limited resolution. (c) Simulated recovered image from the realizable metasurface platform. Pitch and aperture size are assumed to be the same as in Fig. 3b. (d) General schematic representation of angularly-selective light collection with an active metasurface. Images can be reconstructed from arbitrary choice of basis, defined in the incident k-space (schematically shown in darker orange). Measurements $w_n$ give us the weight of basis element n in the image reconstruction. (e) Map of coupling intensity between the incident light and the detector for an example Hadamard basis element used to create Fig. 3f. The weight of each basis element is found through the difference of two measurements, one corresponding to the positive coupling and the other corresponding to the negative coupling. (f) Simulated recovered image with $345 \times 345$ scatterers, $0.14mm \times 0.14\ mm$ ($91\ \lambda \times 91\ \lambda$), imaged with a 4096-element Hadamard basis. Realizable amplitude and phase characteristics are assumed.

In time-critical applications, it may be useful to speed up acquisition time by extracting key information from the scene in significantly fewer measurements by replacing the delta function basis by other k-space bases, as conceptually illustrated in Fig. 3d. Examples of such bases include Hadamard or Fourier bases[26]. In contrast to the point-by-point imaging, however, the realization of these bases generally requires a fully 2D addressable metasurface.

Typically, the lower spatial frequencies of an image are more important than the higher frequencies when it comes to quickly identifying an entire scene. In Fig. 3f, we recover a useful approximation of the scene through an inner product with 4096 basis elements, starting with lower frequency elements[27]. Notably, this image is realized despite a highly non-ideal far field coupling at each measurement (Fig. 3e). Since the Hadamard basis elements have both positive and negative components, we compute each inner product as the difference of two measurements; this difference operation allows an effective negative coupling across portions of the far field. As it was acquired with almost an order of magnitude fewer measurements than the point-by-point retrieval image, this approximate image could be used to select a part of the scene to image in further detail.

The ability of active metasurfaces to generate arbitrary coupling between the far field and the detector additionally indicates that this imaging system may be compatible with compressed sensing. This would enable much shorter imaging times by recovering N-pixel images with only $\sim s \cdot log\left(\frac{N}{s}\right)$ measurements for a $s$ sparse scene[8]. The computational runtime and space complexity of different choices of bases is discussed in SI.8 and are expected to be competitive with other single-pixel and lensless imaging approaches. We additionally demonstrate in SI.9 how reconfigurable coupling can be used to extract information from the scene (e.g., the location of edges) at low computational cost.

**Optical system characteristics**

Next, we consider the proposed imaging system in terms of well-understood metrics: the point spread function (PSF) and the modulation transfer function (MTF). The PSF is the image produced by a single point of light in the far field, and the MTF is the amplitude of the Fourier Transform of the PSF, indicating the maximum modulation depth of the system as a function of spatial frequency. A larger value for the MTF thus reflects the ability to image high frequency information. We compare the PSFs and MTFs of a diffraction-limited pinhole, an ideal active metasurface, and a realizable active metasurface for a fixed aperture size of $51 \: \mu m \times 51 \: \mu m$ ($34\lambda \times 34 \: \lambda$), assuming a point-by-point imaging method. This smaller aperture size is selected such that features of interest remain clearly visible. In this section, we assume negligible noise, achievable with long integration times, such that the results depict only systematic errors.

We find in Fig. 4a and b that the k-space PSF/MTFs achievable by the coupling of an ideal active metasurface with a single detector are identical to those of a pinhole of the same aperture coupled to a far field sensor array. Both PSFs can be approximated as a 2D $sinc^2$ function (see SI.2) and are invariant under translation of the point being imaged when expressed in units of $k_x, k_y$—their corresponding MTFs are thus linear as a function of line-pairs (cycles) per $k_x, k_y$. In contrast, the non-idealities of the realizable active metasurface device produce a non-ideal, angle-dependent PSF. Light from a point source at $0°$ is coupled into each measurement to varying extent—the contribution of the $0°$ source is dependent on the scatterer antenna factor, with the $0°$ source having a larger contribution to measurements at angles where the desired signal couples less efficiently to the designed metasurface. This accounts for the decreased contrast observed in Fig. 3c. Additionally, point sources at oblique angles couple into both the intended $(k_x, k_y)$ measurement and the $(-k_x, -k_y)$ measurement—this explains the 'ghost images' in Fig. 3c. The undesired coupling and resulting imaging aberrations can be fully characterized by computing PSFs at each point of the scene. Because the aberrations do not broaden the PSF but rather introduce predictable, separate peaks, we anticipate that they will be straightforward to correct with computational post-processing. To illustrate this possibility, we consider in Fig. 4c the as-collected image (left) and the two-dimensional $0°$ PSF (center) of the device studied in Fig. 3c. Taking the 2D PSF and weighing it by the known intensity of the far field at $0°$, we can subtract its contribution from the recovered image. The contrast of the recovered image (Fig. 4c) substantially improves from this simple post-processing step. In practical applications, more sophisticated post-processing could also be used to eliminate the 'ghost' image effect. Previous work has also shown that unwanted coupling can be reduced by further array-level inverse design[21], which would improve the unprocessed PSF.

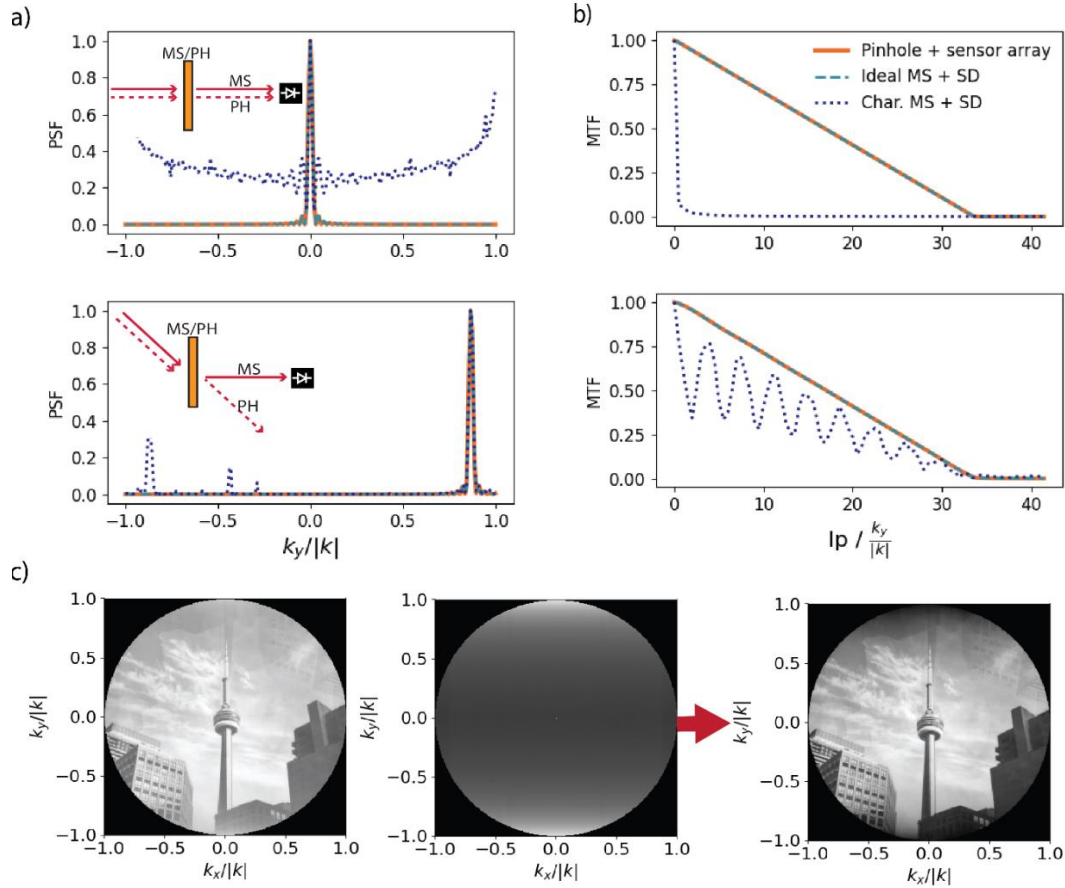

**Figure 4.** Imaging characteristics of a diffraction-limited pinhole camera (solid orange line); of an ideal metasurface with isotropic scattering properties, $2\pi$ phase control, and no amplitude variations (dashed turquoise line); and of an experimentally realizable active metasurface single-pixel imaging device of dipole scatterers (dark blue dotted line). The devices simulated in Fig. 4a and 4b are $51\ \mu m \times 51\ \mu m$ ($34\lambda \times 34\ \lambda$) in size. (a) Point spread function (PSF) along $k_x = 0$ of a point source originating at $(k_x, k_y) = (0,0)$ (top) and $(k_x, k_y) = (0, \frac{\sqrt{3}}{2}|k|)$ (bottom). The insets schematically depict collection of light by an active metasurface (MS) and pinhole (PH). (b) Modulation transfer function (MTF) around the same points as in Fig. 4a. (c) Left: Recovered image from Fig. 3c, showing the effects of the non-idealities of our realizable TCO metasurface. Center: The 2D PSF of the imaging system for a point source at $(0,0)$. This PSF reflects the different dependency of our antenna factor on $k_x$ and $k_y$. Right: Post-processed image where the $(0,0)$ PSF weighted by the $(0,0)$ intensity is subtracted from the recovered image (left) to improve image contrast.

**Detector characteristics in system design**

The stated bounds on the number of resolvable points and the resolution so far assume a diffraction-limited imaging system where the imaging characteristics are dominated by the optical response of the metasurface. However, the effective k-space detector width (angular acceptance range) of the coupled single-detector also affects the PSF of the full system. The k-space detector width is determined by the range of wavevectors around the normal which couple into the detector and is most easily conceptualized for a system with a non-integrated detector element, where this value is determined by the size and distance of the detector to a lens or metasurface.

We consider the impact of detector width, $\Delta k_D$, on the resolution of our imaging system through the PSF/MTF and observe two regimes of operation: a diffraction-limited regime ($\Delta k_D < FHWM_{min}$) and a detector-limited regime ($\Delta k_D > FHWM_{min}$). We describe the assumed detection window in SI.10. For both the ideal and realizable metasurfaces, we find that the PSF and MTFs of the imaging system have little dependence on $\Delta k_D$ in the diffraction-limited regime where $\Delta k_D < FWHM_{min}$ (Fig. 5a, solid lines). However, for $\Delta k_D \geq FWHM_{min}$, shown as dotted lines in Fig. 5a, the PSF broadens proportionally to $\Delta k_D$, resulting in a diminished ability to resolve higher frequency features (Fig. 5a, MTFs). This behavior is expected, as the system PSF is the convolution of the metasurface PSF and the detection window[28]. Then, the number of resolvable points decreases proportionally to $\frac{1}{(\Delta k_D)^2}$ (Fig. 5b).

For a fixed aperture size, however, the rate of photon arrival at the detector increases proportionally to $(\Delta k_D)^2$. Thus, an increased detector width can improve the SNR of an image at the expense of resolution. This also indicates that detector coupling width should never be designed to less than $FWHM_{min}$, as it results in a decrease in efficiency with no gain in resolution. Whereas improving the SNR by reducing the number of measurements through choice of basis requires a fully 2D-addressable metasurface, this approach remains compatible with a perimeter-controlled architecture.

Assuming again a shot-noise limited system, we see as expected that the SNR is improved by increased numbers of ideally collected photons and higher metasurface scattering efficiencies. Thus, the shot-noise limited SNR, shown as black lines in Fig. 5b for different integration times, increases with a wider collection angle. We find that for a fixed imaging time, $SNR \propto (\Delta k_D)^4$ since the time available for each measurement is inversely proportional to the number of measurements, and the photon count collected per unit time scales with detector area. However, in addition to reducing the resolution, increased detector width may introduce a far field dependent normalization error in cases where the antenna factor amplitude is not isotropic. This error arises because we cannot in this case determine the exact coupling loss associated with measurements and is further explained in SI.11.

The effect of the normalization error on our realizable system is included alongside the shot noise error in the scatterplots of Fig. 5b. Shot noise dominates for small detector widths, but as the detector width increases, the normalization error becomes more significant, limiting the SNR to ~40 dB with our chosen bin sizes and the angular dependency of our coupling. This normalization error will decrease to zero as the coupling becomes isotropic over the FOV of interest.

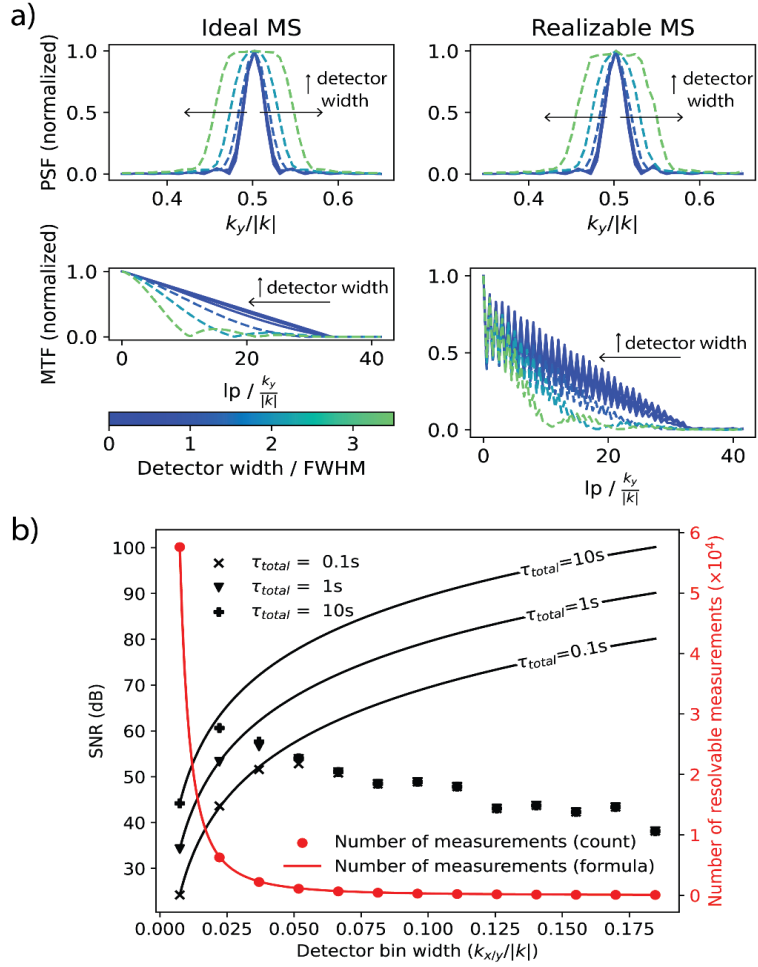

**Figure 5.** Effects of detector bin width (acceptance angle) and integration time on SNR of scenes. Results assume a quasi-monochromatic light source with a photon flux of $7.6 \times 10^{21} \frac{photons}{m^2 \, s}$, an aperture size of $0.2 \, mm \times 0.2 \, mm$, and a single detector with a uniform, square acceptance in k-space. (a) PSF (top) and MTFs (bottom) of ideal (left) and realizable (right) active metasurface single-pixel imager for different detector widths. Solid lines indicate diffraction-limited resolution, whereas dotted lines show detector-limited resolutions. (b) SNR in dB (black, left axis) and number of distinguishable image points (red, right axis) as a function of detector width. The black lines are the shot-noise limited SNR. The black points are the SNR with both shot noise and scene-dependent normalization errors.

## DISCUSSION

We have shown through simulation the potential of coupling active metasurfaces to single-pixel detectors in applications requiring high resolution and wide FOV with compact system dimensions. We provide analytical bounds for information content based on geometric factors of the metasurface and demonstrate a framework which analyzes the effect of losses and modulation rates on acquisition time and SNR. We studied the case of a previously experimentally realized active metasurface and showed that the realizable structure can achieve diffraction-limited images with predictable and thus correctable aberrations. We additionally considered how image acquisition speeds can be improved through dynamically programmable imaging bases or through the design of detector coupling.

These results inform how—depending on the required resolution, SNR, and form-factor of a particular application—different metasurface modulation mechanisms, scatterer designs, and acquisition strategies may be necessary. In general, the following features are desirable to high-speed image acquisition, good resolution, high SNRs, and minimizing aberrations: high efficiency, isotropic antenna factor (coupling) across the FOV of interest, high modulation speeds, $2\pi$ phase control, and large aperture sizes. We refer the reader to other reviews which cover the state-of-the-art on these objectives[9,29,30]. These results also show us tradeoffs of this system. A greater SNR can be achieved by increasing aperture size, thus increasing complexity, or by sacrificing resolution. Fast image acquisition rate requires both high efficiency and fast modulation rates, which are not always mutually realizable.

This analysis allows us to consider design objectives and technological improvements which would be most useful to active metasurface imaging technologies. While the imaging process which we have described is compatible with active metasurfaces in reflection, the integration of a detector element into the metasurface would require a transmissive active metasurface. Critically, such transmissive active metasurface single-pixel imagers could conceivably be built with a very small footprint, making them ideal for applications requiring low SWaP[4]. While dynamically tunable transmissive active metasurfaces designs have been proposed[20], state-of-the-art performance for transmissive active metasurfaces is far less developed than for devices in reflection, leaving significant room for improvement in the available technology.

At present, the manufacture of large (mm)-scale active metasurfaces in two dimensions is a major challenge to achieving high image resolutions[9,31]. While recent fabrication with deep ultraviolet lithography[32] and nanoimprint lithography[33–35] have demonstrated centimeter-scale, high throughput fabrication of passive metasurfaces, 2D addressing of scatterers remains a major obstacle. This is particularly difficult in transmission where the addressing may affect the optical properties of the metasurface. A possible near-term solution to overcome this challenge could be to orthogonally stack pairs of 1D transmissive active metasurfaces or to use a row-column addressing method[22]. While this would not allow a general coupling basis construction, it would be sufficient for point-by-point imaging and considerably easier to fabricate. We discuss state-of-the-art for fully 2D addressing in SI.6.

We also envision several theoretical avenues of investigation. Having demonstrated the flexibility of 2D-addressable active metasurfaces in generating varied coupling bases, a further study into robustness of different bases to noise and fabrication imperfections would be beneficial. We also anticipate that developing metasurface- and basis-specific post-processing algorithms would yield significant improvements. While this work has assumed no phase coherence in time between different points of a scene, we additionally note that active metasurfaces can conserve phase properties of the incident scene. Then, we anticipate that this platform may be able to perform phase-imaging. The best approach and challenges to recovering a full amplitude and phase profile of the scene remains an open question. Finally, while we restricted ourselves to the single-detector case, the formalism from Eqn. 1 can also be used to study small multi-pixel systems, improving acquisition time at the expense of compactness. Extending this work to multiple wavelengths would also allow a richer range of applications.

Finally, this work assumes negligible coupling between adjacent scatterers up to large steering angles. While a well-justified approximation for local (e.g., plasmonic) metasurfaces, in which the mode volume of scatterers is small, coupling is more difficult to avoid in dielectric metasurfaces. Further investigation should be conducted to find how such coupling may affect lensless active metasurface imaging performance.

**MATERIALS AND METHODS**

**Image reconstruction approach**

In this work, we model the imaging process as a series of time-separated measurements, each with some angularly dependent coupling loss and voltage-dependent absorption losses, followed by an image reconstruction algorithm. The image reconstruction algorithms depend on each measurement being normalized, as specified in Eqn. 5.

For computational simplicity, this normalization is implicitly enforced in our image reconstruction as follows:

$$\int_{\frac{\vec{k}_{in}}{|\vec{k}_{in}|} \leq 1} \left|G(\vec{0}, \vec{k}_{in})\right|^2 \left|A(\vec{k}_{in})\right|^2 d\vec{k}_{in} = 1 \qquad (8)$$

Further justification for the normalization and the full image recovery algorithms are provided SI.12 and SI.13.

**Selection of voltage configuration for measurements**

We select the voltage configuration for each target measurement in a few different ways. For the PSF of an ideal metasurface, we create a blazed grating profile where the phase difference between adjacent elements is $d\psi_{x,y} = -k_{x,y}\Delta_{x,y}$, for scatterers arranged with a period $\Delta_{x,y}$ along $x, y$ and target farfield point $(k_x, k_y)$. For point-by-point imaging with a realizable metasurface, we use a single iteration of the GS algorithm to compute an achievable amplitude/phase at each scatterer, then enforce metasurface properties by the method described in SI.14. We found that this approach generates the same far field coupling pattern as selecting an ideal amplitude/phase according to the blazed grating method and then enforcing the metasurface properties, which we use to generate the realizable PSF data. Finally, we use the GS algorithm to generate our Hadamard and kernel far field couplings, using 25 iterations and multiple initializations to improve the achieved coupling.

**Amplitude and phase response of scatterers to voltage**

We calculate the amplitude and phase response of individual scatterers to applied voltage via a Gaussian Radial Basis Function (RBF), fit to the data shown in SI.5. The Gaussian RBF fit is a sum of Gaussian functions, each centered on a voltage data point. We fit the heights of the Gaussians as parameters with Python's implementation of BFGS and maintain a fixed standard deviation of 1.5 V for each Gaussian.

**Data availability:** The code and data used to generate this work is available upon request.

Supplementary information accompanies the manuscript.

**Acknowledgments:**

The authors thank Ruzan Sokhoyan and Jared Sisler for preliminary discussions on the project concept, Claudio Hail for preliminary discussions and suggestions regarding detection methods, Morgan Foley for preliminary discussions and early analysis of 2D imaging, and Lior Michaeli for advice and discussion.

This work was supported by the Meta-Imaging MURI grant #FA9550-21-1-0312 from Air Force Office of Scientific Research. J.B. acknowledges funding through the NSERC PGS D program. P.T. acknowledges support from Meta Platforms, Inc., through the PhD fellowship #C-834952.

**Author contributions:**

P.T. and H.A.A. conceived the original idea. J.B. developed and implemented the simulation formalism, suggested the design of arbitrary imaging bases, performed the imaging studies, and wrote the manuscript. P.T. developed initial 1D imaging calculations and comparisons to conventional imaging methods, investigated possible imaging modes of the proposed system, suggested studies, and revised the manuscript. H.A.A. organized the project, reviewed the formalism, suggested studies, and revised the manuscript. All authors have given approval to the final version of the manuscript.

**Competing interests statement:** The authors declare no conflicts of interest regarding this article.

**Materials & Correspondence:** Harry A. Atwater (haa@caltech.edu)

# Supplementary Information

## 1. Formalism for scattered intensity

We introduce in the manuscript the following equation for the photon flux per unit k-space scattered from the metasurface

$$I_{detected}(\vec{k}_{out}) = \int_{\frac{\vec{k}_{in}}{|k_{in}|}\leq 1} I_{in}(\vec{k}_{in})|G(\vec{k}_{out},\vec{k}_{in})|^2 |A(\vec{k}_{in}-\vec{k}_{out})|^2 d\vec{k}_{in} \qquad (S1)$$

Here, $I_{in}(\vec{k}_{in})$ is the photon flux per unit k-space arriving at the metasurface, $G(\vec{k}_{out},\vec{k}_{in}) = g(\vec{k}_{out})g(\vec{k}_{in})$ describes the complex field coupling between inbound wavevector $\vec{k}_{in}$ and outbound wavevector $\vec{k}_{out}$ for a single metasurface scatterer, with $g(\vec{k})$ being the antenna factor, and where $A(\vec{k}_{in}-\vec{k}_{out})$ is the array factor. Given $\vec{k}_{in}$ and $\vec{k}_{out}$, we can write the antenna factor as

$$A(\vec{k}_{in}-\vec{k}_{out}) = \sum_n a_n e^{i\psi_n} e^{i(\vec{k}_{in}-\vec{k}_{out})\cdot \vec{r}_n} \qquad (S2)$$

This array factor formalism assumes negligible coupling between adjacent elements. We begin by assuming that the Green's function response of the metasurface to a plane wave $\vec{k}_{in}$ can be described as $G(\vec{k}_{out},\vec{k}_{in})A(\vec{k}_{in}-\vec{k}_{out})$, up to normalization, such that the far field amplitudes scattered from the metasurface are

$$E(\vec{k}_{out}) = \int_{\frac{\vec{k}_{in}}{|k_{in}|}\leq 1} E_{in}(\vec{k}_{in}) G(\vec{k}_{out},\vec{k}_{in}) A(\vec{k}_{in}-\vec{k}_{out}) d\vec{k}_{in} \qquad (S3)$$

$$H(\vec{k}_{out}) = \int_{\frac{\vec{k}_{in}}{|k_{in}|}\leq 1} H_{in}(\vec{k}_{in}) G(\vec{k}_{out},\vec{k}_{in}) A(\vec{k}_{in}-\vec{k}_{out}) d\vec{k}_{in} \qquad (S4)$$

Considering integration times $\tau$ greater than $T = \frac{2\pi}{\omega}$, where $\omega$ is the frequency of light. We write the time-averaged Poynting vector as

$$S(\vec{k}_{out}) = \frac{1}{2} E_{detected}(\vec{k}_{out}) \times H^*_{detected}(\vec{k}_{out}) \qquad (S5)$$

$$S(\vec{k}_{out}) = \frac{1}{2}\left(\int_{\frac{\vec{k}_1}{|k_1|}\leq 1} E_{in}(\vec{k}_1) G(\vec{k}_{out},\vec{k}_1) A(\vec{k}_1-\vec{k}_{out}) d\vec{k}_1\right) \times \left(\int_{\frac{\vec{k}_2}{|k_2|}\leq 1} H_{in}(\vec{k}_2) G(\vec{k}_{out},\vec{k}_2) A(\vec{k}_2-\vec{k}_{out}) d\vec{k}_2\right)^* \qquad (S6)$$

$$S(\vec{k}_{out}) = \frac{1}{2}\int_{\frac{\vec{k}_{1,2}}{|k_{1,2}|}\leq 1} [E_{in}(\vec{k}_1)\times H^*_{in}(\vec{k}_2)] G(\vec{k}_{out},\vec{k}_1) G^*(\vec{k}_{out},\vec{k}_2) A(\vec{k}_1-\vec{k}_{out}) A^*(\vec{k}_2-\vec{k}_{out}) d\vec{k}_1 d\vec{k}_2 \qquad (S7)$$

Consider now a time $\tau \gg \frac{2\pi}{\omega}$. This will be true for integration times of interest in imaging. In this case, we can average $S$ over the time $\tau$

$$S_{detected}(\vec{k}_{out}) = \frac{1}{\tau}\int_{t-\tau}^{t} S(\vec{k}_{out},t') dt' \qquad (S8)$$

While the active metasurface modifies a narrowband phase and amplitude response, we assume that the incident light itself is not fully monochromatic, such that it is incoherent in time over the integration time of the detector. Then, assuming that our far field is of constant intensity but time incoherent, we note that the cross product of fields from different far field points average to zero on the timescale of $\tau$. That is,

$$\frac{1}{\tau}\int_{t-\tau}^{t}\frac{1}{2}E_{in}(\vec{k}_1,t')\times H_{in}^*(\vec{k}_2,t')dt' = S_{in}(\vec{k}_1)\delta(\vec{k}_1-\vec{k}_2) \tag{S9}$$

Then, Eqn. S7 reduces to

$$S_{detected}(\vec{k}_{out}) = \int_{\frac{\vec{k}_{in}}{|\vec{k}_{in}|}\leq 1} S_{in}(\vec{k}_{in})|G(\vec{k}_{out},\vec{k}_{in})|^2|A(\vec{k}_{in}-\vec{k}_{out})|^2 d\vec{k}_{in} \tag{S10}$$

Finally, we retrieve Eqn. S1 by substituting the Poynting vector per unit k-space by the photon flux per unit k-space, which is proportional to the Poynting vector.

## 2. Analytical bounds on resolution and number of resolvable points in a scene

In this section, we derive the k-space full width half max (FWHM) achievable by an ideal active metasurface as a measure of achievable image resolution. We also obtain an analytic expression for the number of points resolvable by an active metasurface with subwavelength scatterers.

We study the case of point-by-point imaging to bound system performance. To achieve a diffraction-limited resolution, we set a constant phase gradient such that $\psi_n^{x/y} = \Delta\psi_{x/y} n_{x/y}$ where light is collected from target wavevector $\vec{k}_t = (k_{tx}, k_{ty})$ by setting phase gradients $\frac{\Delta\psi_x}{\Delta_x} = -2\pi k_{tx}$, $\frac{\Delta\psi_y}{\Delta_y} = -2\pi k_{ty}$. Here, the quantities $\Delta_x, \Delta_y$ are the metasurface pitch along the x and y directions respectively. Then, the array factor of the system becomes (Eqn. S2)

$$A(\vec{k}_{in}) = \sum_{n_x,n_y} e^{i(\Delta\psi_x n_x + \Delta\psi_y n_y)} e^{i(k_x^{in}\Delta_x n_x + k_y^{in}\Delta_y n_y)} \tag{S11}$$

For notational simplicity, we define $\gamma_i = \Delta\psi_i + k_i^{in}\Delta_i$, $i \in \{x,y\}$, and write

$$A(\vec{k}_{in}) = \sum_{n_x} e^{i\gamma_x n_x} \sum_{n_y} e^{i\gamma_y n_y} = \left(\frac{1-e^{i\gamma_x N_x}}{1-e^{i\gamma_x}}\right)\left(\frac{1-e^{i\gamma_y N_y}}{1-e^{i\gamma_y}}\right) \tag{S12}$$

$$|A(\vec{k}_{in})|^2 = \left(\frac{\cos(\gamma_x N_x)-1}{\cos(\gamma_x)-1}\right)\left(\frac{\cos(\gamma_y N_y)-1}{\cos(\gamma_y)-1}\right) \tag{S13}$$

Next, we assume that $N_x, N_y \gg 1$. This allows us to make two statements:

1. The FWHM of our coupling is dominated by $|A(\vec{k}_{in})|^2$ rather than $|G(\vec{0},\vec{k}_{in})|^2$.
2. We can Taylor expand $(\cos(\gamma_i)-1)^{-1}$ around $\gamma_i = 0$, as at large $N_i$, the FWHM occurs at $\gamma_i \ll 1$.

$$|A(\vec{k}_{in})|^2 = N_x^2 N_y^2 \text{sinc}^2\left(\frac{\gamma_x N_x}{2\pi}\right)\text{sinc}^2\left(\frac{\gamma_y N_y}{2\pi}\right), \quad \gamma_i \ll 1 \tag{S14}$$

where we define $sinc(x) = \frac{\sin(\pi x)}{\pi x}$.

This function has a maximum of $N_x^2 N_y^2$ at $\gamma_x = \gamma_y = 0$. Thus, the coupling intensity drops by half along each axis when

$$sinc(H) = \frac{1}{\sqrt{2}}, \quad H = \frac{\gamma_i N_i}{2\pi} \approx 0.443 \tag{S15}$$

Then, our FWHM normalized by $|k|$ can be solved as

$$FWHM_i = \frac{2H\lambda}{N_i \Delta_i} \tag{S16}$$

Thus, a single resolvable point occupies an area of $FWHM_x \cdot FWHM_y$ in k-space. Finally, we can divide the total k-space area available to our imaging system by the area of a single resolvable point to find the number of distinct points, $N_p$, which can be resolved by the system. This total available area depends on the numerical aperture $NA$, or equivalently on the field of view (FOV), accessible to the metasurface.

$$N_p = \frac{\pi \cdot NA^2}{4H^2} \left(\frac{N_x \Delta_x}{\lambda}\right)\left(\frac{N_y \Delta_y}{\lambda}\right) \tag{S17}$$

## 3. Fundamental limits on the field of view of imaging

This analysis and discussion follows the work of Kim et al.[1]. As in the main text, we assume that the metasurface scatterers are distributed on a rectangular grid. We also assume that the grid is subwavelength, that is, that $\frac{\lambda}{\Delta_x} \geq 1, \frac{\lambda}{\Delta_y} \geq 1$. There are two types of FOV which we consider in this analysis. Though generally, the regions in k-space where aliasing-free imaging is achievable are not circular (i.e., do not directly correspond to a FOV), we report here on the greatest FOV which falls fully within the aliasing-free region, for simplicity. We refer the reader to Kim et al.'s work for further analysis on the shape of alias-free regions and the effect of lattice shape[1].

We first introduce a 'strong FOV,' in which light can be coupled into the normal without undesired higher coupling orders existing within the light cone. This means that alias-free imaging can be achieved across this FOV given any antenna factor (called the 'weighting factor' in Ref.1) which is non-zero across the full strong FOV. More precisely, it denotes the FOV corresponding to the largest NA that falls fully within the aliasing-free regime[1].

We can begin our analysis with the same array factor derived in SI.2

$$\left|A(\vec{k}_{in})\right|^2 = \left(\frac{\cos(\gamma_x N_x) - 1}{\cos(\gamma_x) - 1}\right)\left(\frac{\cos(\gamma_y N_y) - 1}{\cos(\gamma_y) - 1}\right) \tag{S13}$$

This coupling is maximized under the following condition

$$\gamma_x = \Delta\psi_x + k_x^{in}\Delta_x = 2\pi m_x \tag{S18a}$$

$$\gamma_y = \Delta\psi_y + k_y^{in}\Delta_y = 2\pi m_y \tag{S18b}$$

where $m_i$ are integers corresponding to the diffracted order number along the $i$ axis and we assume that $\Delta\psi_x, \Delta\psi_x \in [-\pi, \pi)$.

We re-express the condition in terms of the target steering in unit k-space. Let $(u_x, u_y) = \frac{\lambda}{2\pi}(k_x, k_y)$. Note that to be in the light cone, it must be true that $u_x^2 + u_y^2 \leq 1$. Then, we can write the unit k-space points at which diffracted orders appear as

$$u_x = -\left(\frac{\Delta\psi_x}{2\pi} - m_x\right) \cdot \frac{\lambda}{\Delta_x} = -u_{x0} + m_x \frac{\lambda}{\Delta_x} \tag{S19a}$$

$$u_y = -\left(\frac{\Delta\psi_y}{2\pi} - m_y\right) \cdot \frac{\lambda}{\Delta_y} = -u_{y0} + m_y \frac{\lambda}{\Delta_y} \tag{S19b}$$

where $u_{i0} = \frac{\Delta\psi_i}{2\pi}\frac{\lambda}{\Delta_i}$ describe the zeroth order diffraction peak position along axis $i$.

The $m_x^{th}, m_y^{th}$ diffracted orders can exist in the unit k-space domains defined below

$$D_{mx,my} = \left\{(u_x, u_y) \,\Big|\, \left(u_{xo} - m_x \frac{\lambda}{\Delta_x}\right)^2 + \left(u_{yo} - m_y \frac{\lambda}{\Delta_y}\right)^2 \leq 1\right\} \tag{S20}$$

where each domain $D_{mx,my}$ is a circle in unit k-space, with unity radius, and centered at $\left(m_x \frac{\lambda}{\Delta_x}, m_y \frac{\lambda}{\Delta_y}\right)$. The points in unit k-space which are accessible in the aliasing-free regime are those which fall within the domain $D_{00}$ but not any higher order domain, that is

$$D_{AF} = D_{00} - D_{mx,my} \,\forall\, m_x \neq 0 \text{ or } m_y \neq 0 \tag{S21}$$

It can be shown for a metasurface with scatterers positioned along a rectangular grid that the NA which falls fully within the aliasing-free zone is limited by transverse wave vectors which are fully oriented along the x or y axes[1]. This maximum can be written as

$$NA_s = u_x^{max,AF} = \min\left(\frac{\lambda}{\Delta_x} - 1, \frac{\lambda}{\Delta_y} - 1\right) \tag{S22}$$

Thus

$$FOV_s = 2\arcsin\left(\min\left(\frac{\lambda}{\Delta_x} - 1, \frac{\lambda}{\Delta_y} - 1\right)\right) \tag{S23}$$

Whether through mechanical filtering of light or through scatterer design, it is generally possible to prevent the coupling of light incoming from an angle greater than a threshold value. This provides us with a 'weak FOV' which falls within the Brillouin zone, in which the undesired higher orders have greater transverse components than the desired coupling order and can thus be filtered out[1]. To get an upper bound on achievable FOV, we thus assume that each scatterer has a designed antenna factor such that the antenna factor is significant within the BZ, and negligible outside of it. This 'weak FOV' then denotes the maximum FOV which would be achievable given such an ideally tailored antenna factor[1]. In the limit of this antenna factor, the higher diffraction orders which appear in the light cone but remain outside of the BZ do not interfere with our measurements. The boundary of the BZ is determined by the condition $|\Delta\psi_x| = |\Delta\psi_y| = \pi$ or equivalently $|u_{i0}| = \frac{\lambda}{2\Delta_i}$. We thus define our weak NA and FOV as

$$NA_w = \min\left(\frac{\lambda}{2\Delta_x}, \frac{\lambda}{2\Delta_y}\right) \tag{S24}$$

$$FOV_w = 2\arcsin\left(\min\left(\frac{\lambda}{2\Delta_x}, \frac{\lambda}{2\Delta_y}\right)\right) \tag{S25}$$

The listed bounds reflect the direction in which the imaging system is most limited in FOV. In practice, there are certain directions in which non-aliased measurements could be taken beyond the stated angles[1]. It is also important to note that light collection from within the weak FOV may still suffer from lower efficiency than in the strong FOV. In the SNR discussion in the main text, it is assumed that $\frac{\lambda}{\Delta x}, \frac{\lambda}{\Delta y} \geq 2$ and thus that every imaged point falls within the strong FOV ($FOV_s = 180°$). However, the analysis remains applicable across a smaller strong FOV.

## 4. Shot-noise dominated SNR derivation

We introduce the following formulation for signal-to-noise ratio (SNR) of the image in the manuscript

$$SNR = \frac{\sum_{k_x,k_y} N_{ideal}[k_x, k_y]^2}{\sum_{k_x,k_y} \eta^{-1}[k_x, k_y] N_{ideal}[k_x, k_y]} \tag{S26}$$

We derive the following from the spatial domain SNR described by Gonzalez and Woods in Chapter 5 of Digital Image Processing[2]

$$SNR = \frac{\sum_{k_x,k_y} S[k_x, k_y]^2}{\sum_{k_x,k_y} E[k_x, k_y]^2} \tag{S27}$$

where $S$ is the imaged intensity and $E$ is the error in intensity at each measurement. We assume that shot noise dominates the noise in the system to quantify the impact of the active metasurface on the collected image SNR, noting that more generally we would expect specific detectors to contribute to noise in their own way. The shot noise depends on number of photons detected, $N_{detected}$, which can be derived from the rate of photon arrival $P$ ($photons/s$) at the detector, quantum efficiency $Q_e$, and measurement integration time $\tau$. Since the standard deviation of shot noise $\sigma_{shot}$ goes like the square root of the average number of events, we can write

$$\sigma_{shot} = \sqrt{N_{detected}} \tag{S28}$$

This value depends on the far field, the efficiency with which light couples into the metasurface, and the metasurface losses, which vary with the metasurface configuration.

We note that the image is reconstructed from the ideal photon count, $N_{ideal}$, as specified in the manuscript, which is obtained by rescaling the detected photon count.

$$N_{ideal}[\vec{k}_t] = \left(\eta_{\vec{k}_t}\right)^{-1} N_{detected}[\vec{k}_t] \tag{S29}$$

Critically, the error introduced by shot noise is also scaled. Then, we can re-express Eqn. S27 as:

$$SNR = \frac{\sum_{k_x,k_y} N_{ideal}[k_x, k_y]^2}{\sum_{k_x,k_y} \left((\eta[k_x, k_y])^{-1} \sqrt{N_{detected}[k_x, k_y]}\right)^2} \tag{S30}$$

$$SNR = \frac{\sum_{k_x,k_y} N_{ideal}[k_x,k_y]^2}{\sum_{k_x,k_y}\left((\eta[k_x,k_y])^{-1}\sqrt{\eta[k_x,k_y]N_{ideal}[\vec{k}_t]}\right)^2} \tag{S31}$$

We see that we obtain Eqn. S26 by simplifying the denominator. Finally, we get the SNR in dB as

$$SNR_{dB} = 10\log_{10}(SNR) \tag{S32}$$

## 5. Assumed active metasurface characteristics

Our simulation of the 'realizable' active metasurface assumes that the amplitude and phase response of individual scatterers follows that of the experimentally demonstrated TCO-based plasmonic metasurface presented in Ref. 3 (see Fig. S1). The dots correspond to data collected from full-wave simulations of the metasurface at $\lambda = 1509.8\ nm$, which were found to be in good agreement with experimental performance[3]. The line is the Gaussian Radial Basis Function (RBF) fit used in our simulations.

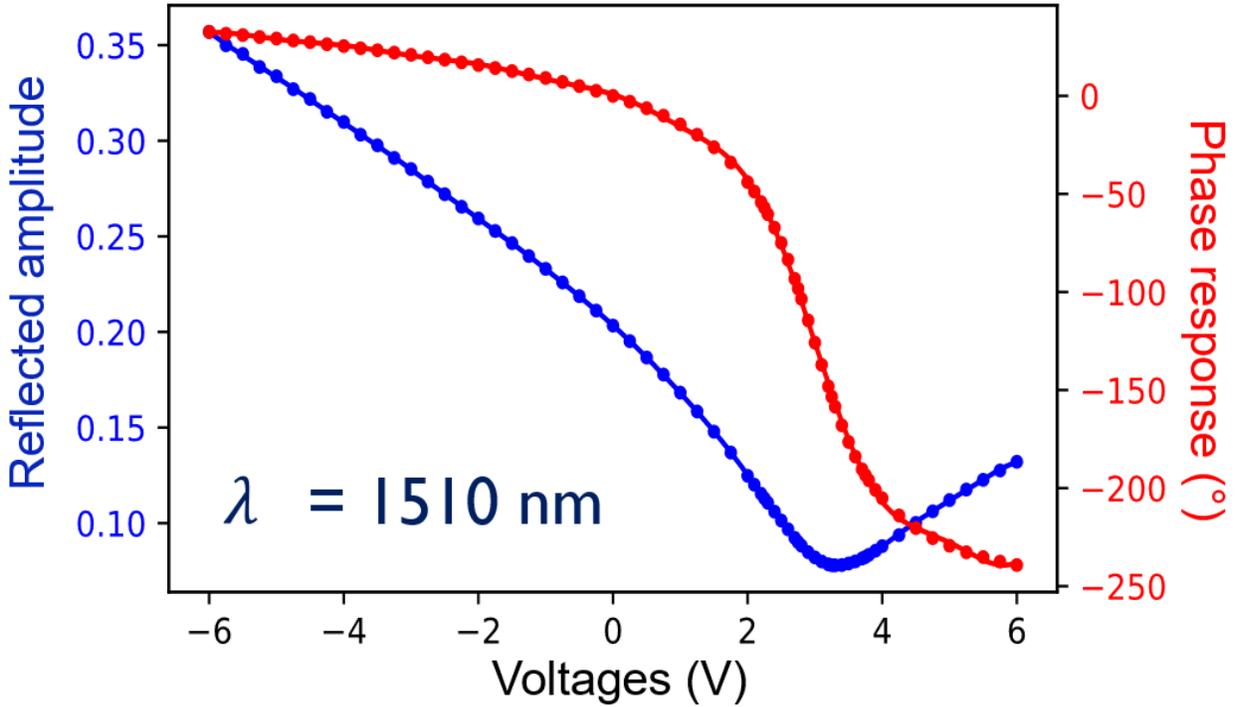

**Figure S1.** Amplitude and phase of scattered light as a function of applied voltage on a TCO-based plasmonic metasurface, at $\lambda = 1509.8\ nm$. The dots correspond to data obtained through full-wave simulations and validated against experiment. The lines correspond to a Gaussian RBF best fit.

## 6. Two-dimensional addressing of active metasurfaces

By using a perimeter-control architecture[4], the number of biasing lines required to address an active metasurface can be reduced from $N^2$ to $2N$, resulting in a significantly simpler design and fabrication process. This is a commonly used approach in the production of commercial spatial light modulators[5]. We can consider two ways of utilizing such an architecture.

The first approach consists of applying 2N signals to set the metasurface configuration. This does not allow full 2D control of the active metasurface, but rather enables the creation of phase/amplitude masks which are the outer product between two linearly independent vectors[4]. This is sufficient for beam-steering applications and thus for point-by-point imaging, but not for arbitrary coupling bases such as Hadamard and Fourier bases. It is advantageous, however, due to the high frequency at which the full metasurface can be reconfigured.

Alternatively, perimeter control can achieve fully two-dimensional control with the addition of capacitive elements which store a scatterer state. This has been demonstrated in reflection down to a pixel size of $1\mu m \times 1\mu m$ in a $480 \times 640$ pixel array[6]. The necessity of scanning across the full array, however, imposes additional constraints on the achievable modulation rates. Alternatively, in the infrared, two-dimensional wavefront control has been achieved by designing individual biasing lines for each element[7]. However, the demonstrated device size was small, and such a technology may be difficult to scale with aperture size as described in this report. The design of control architectures will thus be a major challenge in the development of commercially viable devices. We refer to a previous review for a more thorough analysis of possible addressing methods[5], and leave additional study of scalability to future work.

## 7. Comments on feasibility of wide field of view imaging

In our work, we use the array factor formalism to demonstrate the possibility of wide FOV imaging. In the formalism, it is assumed that the amplitude and phase response of scatterers is constant with the angle of the incoming plane wave, and that said response matches the experimental responses calculated at normal incidence.

While we expect experimental variation in scatterer responses as a function of angle, we note that by reciprocity, our metasurfaces should be able to gather light from a wide angle as efficiently as they can steer light to wide angles. Thus, if we aim to collect light towards normal incidence, it is reasonable to consider the scatterer responses to normal incidence. We verify this intuition through full-wave simulations of a plasmonic stripe antenna metasurface[8] (Fig. S2b) based on Ref. 8, performed with Lumerical FDTD. This simulation confirms that we expect strong beam steering, and conversely light collection, performance up to an angle of 71° with the metasurface design considered (Fig. S2a).

Generally, the metasurface pitch, antenna factor (single scatterer behavior) and inter-element coupling all must be considered when evaluating the FOV over which a metasurface can efficiently collect and steer light. Inter-element coupling can significantly reduce the diffraction efficiency of a metasurface, particularly when steering to wide angles. Our analysis assumed minimal inter-element coupling, as tends to be appropriate for plasmonic metasurfaces—we leave the analysis of the impacts of inter-element coupling for future work.

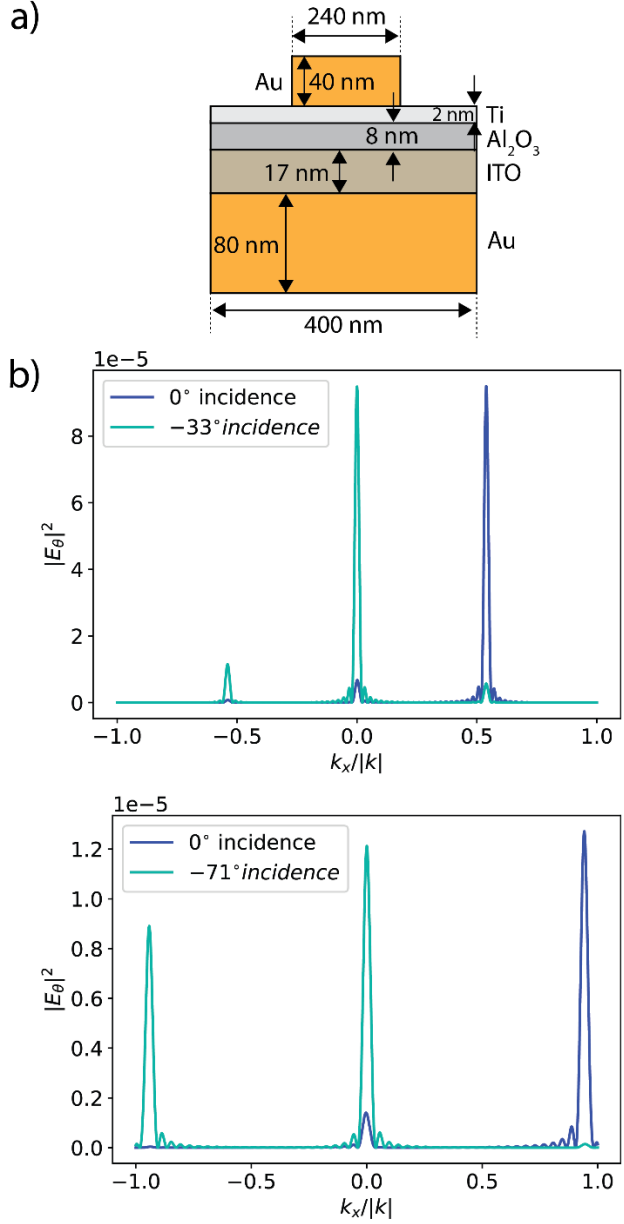

**Figure S2.** FDTD demonstrations of metasurface beam steering reciprocity for $\lambda = 1510\ nm$. The metasurface is simulated with periodic boundary conditions around each unit cell.

(a) Simulated scatterer. The total simulated array size consists of 96 of these elements.

(b) Steering simulated with periodic boundary conditions. *Top*: Steering from 7-element unit cell, where the phase difference $\Delta\psi$ between adjacent elements is set to $\Delta\psi = \frac{2\pi}{7}$. This steers the beam to/from a polar angle $\theta = 33°(k_x = 0.54k)$. *Bottom*: Steering from 4-element unit cell, where the phase difference $\Delta\psi$ between adjacent elements is set to $\Delta\psi = \frac{\pi}{2}$, which steers the beam to/from a polar angle $\theta = 71°(k_x = 0.94k)$.
8

## 8. Computational runtime and memory costs for image recovery

In this section, we will take $N$ to be the number of pixels/datapoints which we aim to retrieve, $s$ to be the sparsity of the scene for some known basis, and $M$ to be the number of independently addressed scatterers. Note that these asymptotic limits include only the cost of processing the acquired data, and not of data acquisition itself. Data acquisition is briefly discussed at the end of this section of the SI.

Consider first a point-by-point imaging approach. Each one of $N$ measurement provides us with a scalar value which must be renormalized by a known (previously characterized) efficiency value. Thus, the computational time for the recovery of the image is $O(N)$. The total space complexity of processing the image is also $O(N)$, and is required to store the image data.

Next, consider a method based on orthogonal bases. The Hadamard basis shown in Fig. 3e is recovered in a total of $N$ measurements, which provide a final image dimension of $\sqrt{N} \times \sqrt{N}$ [9]. Each measurement provides the inner product between the scene and a basis element of size $N$. Thus, the computational cost of the image recovery is $O(N^2)$. While this cost is higher than for the point-by-point imaging approach, we expect that an understanding of the scene can be generated with smaller $N$ with a Hadamard basis. Moreover, the addition of each new basis element can be done in parallel with acquisition, further reducing the overall time required to produce an image. As we can add the contribution of each basis element to a single array of image values, the total memory cost remains $O(N)$. This assumes that individual basis elements can be generated in $O(N)$ and do not need to be stored, which is the case for the Hadamard basis.

We can also consider image retrieval with a k-space Fourier basis[9]. To recover an image with this basis, we find each Fourier spectrum component with 3 or 4 distinct measurements. This must be done $N$ times and as such, we can retrieve the full Fourier spectrum of our image in $O(N)$. We can then apply an inverse Fourier transform to reconstruct the image using the Fast Fourier transform (FFT), which has a runtime of $O(N \log N)$. The overall runtime is thus $O(N \log N)$. The FFT can also be performed in place, maintaining a total space complexity of $O(N)$.

For an analysis of the computational complexity of compressed sensing approaches, we refer the readers to a previous review[10], as many different reconstruction approaches exist.

Finally, it is important to note that an additional space complexity of $O(MN)$ is introduced if we additionally consider the task of setting the metasurface voltage configuration. For our point-by-point imaging demonstration in Fig. 3b and 3c, assuming that 16 discrete voltage levels are available to set at each element (4 bits), this requires $MN \times 0.5 \, Bytes = 511^2 \times 57609 \times \frac{1}{2} \, Bytes = 7.5 \, GB$ of memory for a fully 2D addressable aperture and $MN \times 0.5 \, Bytes = 2 \times 511 \times 57609 \times \frac{1}{2} \, Bytes = 29 \, MB$ of memory for a perimeter-addressed platform. By the same calculation, implementing the Hadamard basis example would require $0.24 \, GB$. It is also generally possible to generate the metasurface configuration in place, but we would not recommend it because it would significantly increase the computational runtime of imaging.

## 9. Single-pixel edge detection and computational edge detection

We use the reconfigurable array factor of the active metasurface to extract information from the scene (e.g., the location of edges) at low computational cost. Once again, a fully 2D addressable metasurface is assumed to retrieve these results. We focus on edge-detection due to its significance in machine vision applications[11]. We use Fig. S3a as our choice of scene because it has distinct edges of varying scale and orientations. The full image covers a 180° FOV, however, we aim to detect edges only out to the shaded line (128° FOV) because we need spatial information around a point to detect an edge, which we cannot have at the boundary of the scene.

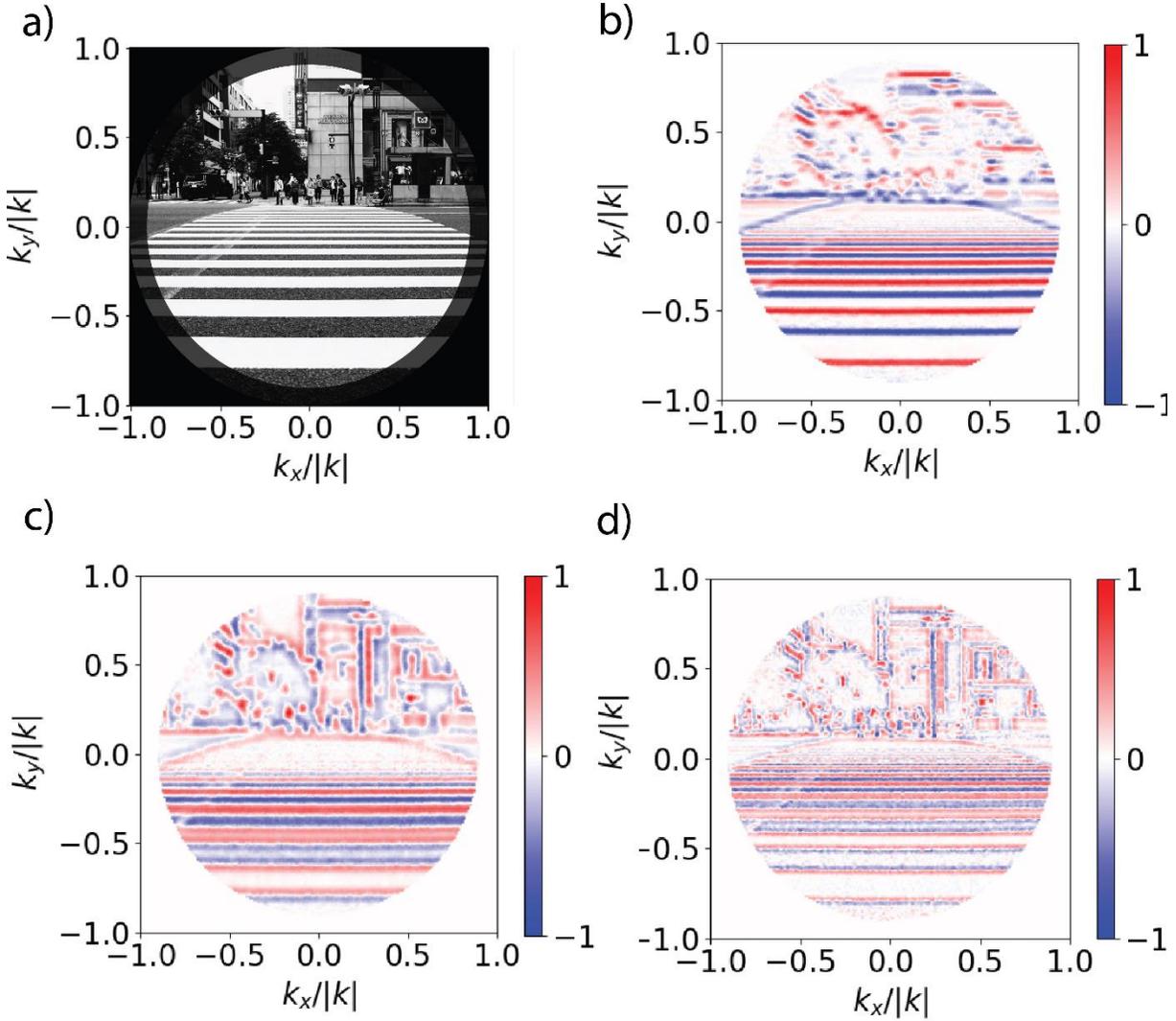

**Figure S3.** Simulated single-pixel edge-detection via an active metasurface imaging device. All results were generated assuming an active metasurface with $345 \times 345$ scatterers, $0.14\ mm \times 0.14\ mm$ ($91\ \lambda \times 91\ \lambda$).

(a) Quasi-monochromatic (single wavelength) ground truth scene. The bright circle depicts a 128° FOV and is the part of the scene over which edge detection is performed. The dimmer circle extends to a 180° FOV. Picture by Daryan Shamkhali on Unsplash.

(b-d) Simulated edges of the ground truth scene detected by each type of optimized kernel. Direction-sensitive edge detection with a Sobel kernel (b), edge detection of large features with a large Laplacian of Gaussian kernel (c), edge detection of small features with a small Laplacian of Gaussian kernel (d).

Edges of an image are commonly found by the computational convolution of an initial image and various kernels, which can be computationally expensive on large images. Here, we instead perform our convolutions by selecting metasurface configurations which weigh the coupling of light into the detector by edge-detection kernels. We find that this method allows us to perform various types of edge detection (Fig. S3b-d).

To properly compare our proposed single-pixel edge detection approach to common methods of computational edge detection, it is useful to consider a standard workflow for edge detection (Fig. S4b). In computational edge detection, an initial image is acquired and converted to a digital array of intensities. It may also be downscaled, to speed up the edge detection process at the expense of some loss of information. The image is then blurred, enough to reduce noise but not so much that it eliminates relevant features, and convolved with edge detection kernels. In each convolution, an $M \times M$ kernel matrix is pointwise multiplied with an $M \times M$ block of our $N \times N$ image. The image block being multiplied with the kernel matrix is shifted across the full image with some stride $S$ (i.e., the pointwise matrix multiplication is taken every $S$ pixels). Finally, information from different kernels is recombined as necessary, here by a sum of squares and thresholding. The overall runtime of the convolution is $O(N^2 M^2 / S^2)$.

For large images and kernel sizes, this process becomes computationally expensive due to the repeated pointwise multiplication of matrix elements. For this reason, it is common for images to be downscaled and for kernel sizes to be limited to 3 x 3 or 5 x 5 pixels, though in recent years there has been renewed interest in large kernel designs in convolution neural networks (CNNs)[12]. To identify larger image features with small kernels, it becomes necessary to either downscale the initial image until image features are on the scale of ~5 x 5 pixels, or more commonly, to build in sequential layers of convolution as in CNNs. In Fig. S4b, we choose to use 7 x 7 Sobel kernels for edge detection, though in CNNs, kernels are often trained and optimized rather than pre-determined.

Active metasurfaces offer an alternate way of taking the convolution of an image with a kernel, compatible with single-pixel detection. We can obtain the convolution of the scene with an arbitrary kernel at a point by optimizing the array factor to first the positive, then the negative component of said kernel, centered at each point of interest. The first measurement then gives us the convolution of the positive portion of our edge-detection kernel with the scene, up to some known rescaling, and the second measurement does the same for the negative portion of our edge-detection kernel. The scalar difference of these two measurements then gives us the measurement at the point, without requiring pointwise multiplication of matrices. We show the results of this process in Fig. S4d (duplicated from Fig. S3b-d). As in our Hadamard image reconstruction, we note that imperfect kernels (Fig. S4c) nevertheless result in clear edge detection. From left to right, we demonstrate directional (Sobel) edge detection, large-feature edge detection (large Laplace of Gaussian), and small-feature edge detection (small Laplace of Gaussian) to showcase the versatility of our approach. In these figures, we evaluate the kernels at 21 265 distinct points (42 530 measurements), but we could reduce the number of measurements by increasing the stride (spacing) between adjacent measurements as is often done in computational edge-detection.

Thus, to perform edge detection, we can either acquire a full image to a desired resolution and go through the computational steps described above, with a relatively computationally expensive pointwise matrix multiplication step associated with each image point, or at most double the number of measurements taken (depending on choice of stride) and take the difference of two scalars at the same image points. We expect this second approach to be particularly valuable for spatially large convolution kernels, as the acquisition time can be decoupled from choice of kernel (though losses may vary in a practical system), unlike the computational cost of the matrix multiplication. More generally, these results reinforce that the variety of measurements which can be achieved with single-pixel active metasurface devices allows for information to be extracted from the scene without full scene reconstruction. We would recommend that future studies look further into potential applications of the technology in compressive classification and in the design of "smashed" (dimensionally reduced) filters which collect only necessary information for classification[13].

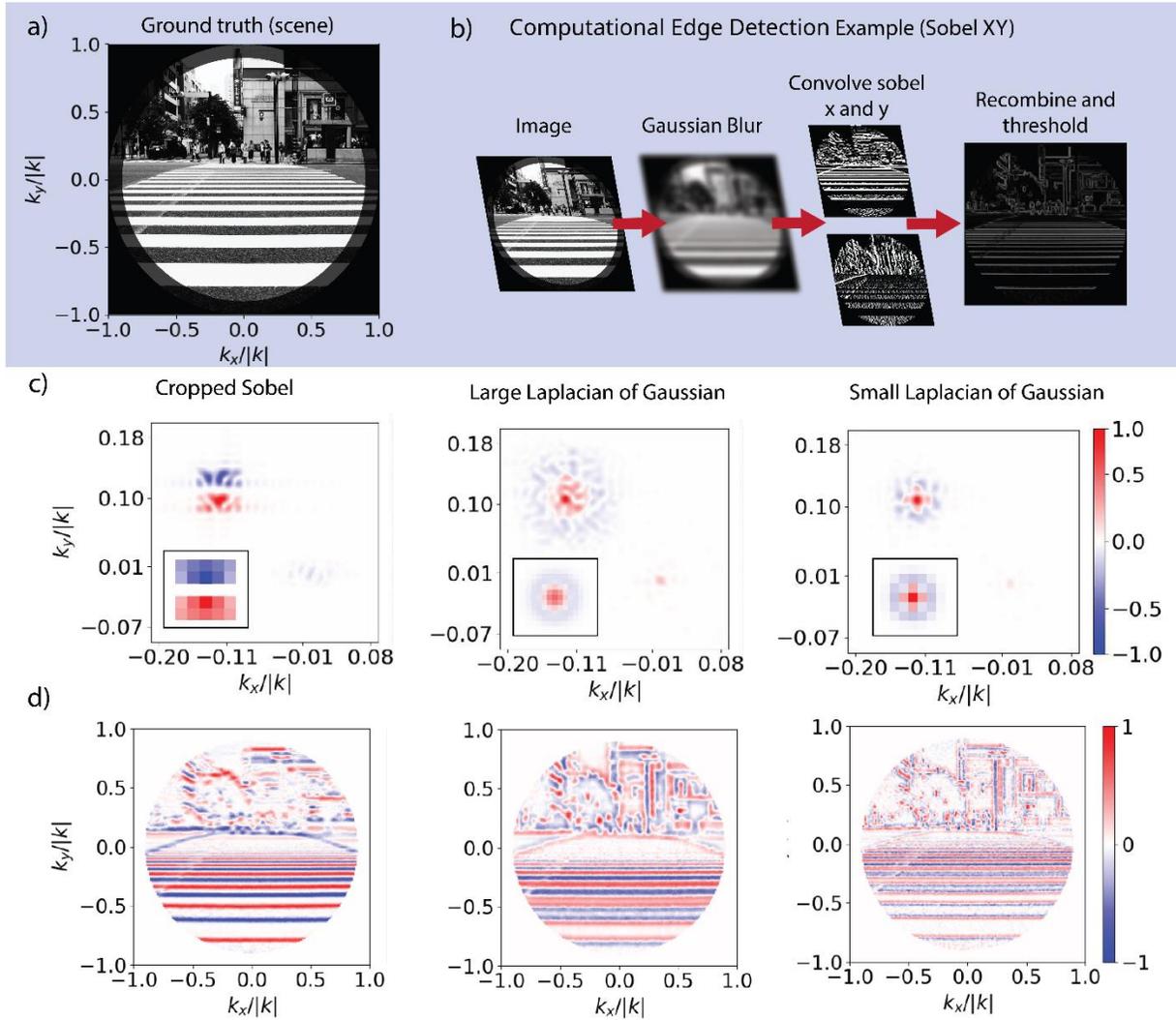

**Figure S4.** Simulated single-pixel edge-detection via an active metasurface imaging device. All results were generated assuming an active metasurface with $345 \times 345$ scatterers, $0.14mm \times 0.14\ mm$ ($91\ \lambda \times 91\ \lambda$).

(a) Quasi-monochromatic (single wavelength) ground truth scene, replicated from Fig. S3a. The bright circle depicts a 128° FOV and is the part of the scene over which edge detection is performed. The dimmer circle extends to a 180° FOV. Picture by Daryan Shamkhali on Unsplash.

(b) A standard workflow for computational edge detection in machine vision applications.

(c) Map of computed coupling intensities as a function of $k_x, k_y$ for a pair of measurements. The difference between the two measurements is the product of the farfield scene with the coupling. The sobel kernel (left) approximates a first derivative in $k_y$. The Laplacian of Gaussians approximate second order derivatives at a large (center) and small (right) scale. Insets represent the target kernels provided to the G-S optimization.

(d) Simulated edges of the ground truth scene detected by each type of optimized kernel, replicated from Fig. S3b-d. Left to right: Direction-sensitive edge detection, edge detection of large features, edge detection of small features.

## 10. Assumptions for k-space bin width

To simplify our analysis, we assume a square detector in k-space with a width of $\Delta k_D$ such that the detection efficiency $D$ as a function of wavevector $(k_x, k_y)$ is

$$D(k_x, k_y) = Q_e\, \Theta\left(\frac{\Delta k_D}{2} - |k_x|\right) \Theta\left(\frac{\Delta k_D}{2} - |k_y|\right) \tag{S33}$$

where $\Theta$ is the Heaviside step function and $Q_e$ is the quantum efficiency, the ratio of incident photons to collected charge carriers. We generally expect that the trends observed in Fig. 5a of the manuscript to hold for other detector shapes as well.

We calculate the number of resolvable points by dividing the total k-space by the width of a detection bin in k-space. In Fig. 5b of the manuscript, this calculation assumes that we can image a FOV of 180°. The red line is computed directly from the ratio of areas. The red dots are a count of the number of measurements which fully fit our FOV and take into account the fact that as bin width increases, discretization effects make the number of available measurements differ from the ratio of areas. We note that the number of resolvable points $N_p \propto \frac{1}{(\Delta k_d)^2}$.

## 11. SNR calculations with varying bin widths

Figure 5b of the manuscript accounts for two sources of errors—the shot noise, as described in Eqn. S28 and a systematic error due to normalization. This error can best be understood through an example. Suppose that an active metasurface has a dipole antenna factor, that the metasurface configuration is set such that each scatterer has the same amplitude and phase response, and that we have a relatively wide detector width $\Delta k_D/|k| = 0.2$. Assume also for simplicity that the scattering efficiency is only lowered by imperfect coupling at wider angles. Then, the detector will collect light from the normal, $(k_x, k_y) = (0,0)$, with an efficiency of $\eta = 1$, as well as from other angles near the normal, including $(k_x, k_y) = (0, 0.1)$ with an efficiency of $\eta = 0.99$. Not knowing *a priori* where the light is coming from, we cannot distinguish in our measurement the effect of $0.99N$ photons arriving from the normal from $N$ photons arriving from $(k_x, k_y) = (0, 0.1)$. That is, because we cannot distinguish receiving fewer photons with higher coupling efficiency from receiving more photons with lower coupling efficiency, the correct choice of normalization lies in $\eta \in (0.99, 1)$ and we do not know what it is. The best we can do is to take the average required normalization across the detector. This results in a systematic error which depends on the scene being imaged, and which is most pronounced when the antenna factor changes rapidly over the width of the detector. Thus, with a wider detector width, this error increases.

We formalize our treatment of this error below. Consider a measurement made with a detection bin wider than the diffraction limit of the active metasurface. We start our calculation from an image, $N_{ideal}[k_x, k_y]$, where $k_x, k_y$ values are spaced within the diffraction limit. A non-diffraction-limited measurement $N_{detected,nd}[\widetilde{k_x}, \widetilde{k_y}]$ gathers a photon count which can be approximated as

$$N_{detected,nd}[\widetilde{k_x}, \widetilde{k_y}] = \sum_{k_i \in \widetilde{k_i} \pm \Delta k_d} N_{detected}[k_x, k_y] \tag{S34}$$

Here, $\widetilde{k_x}, \widetilde{k_y}$ indicate the center of our non-diffraction limited measurements, which are more sparsely distributed than the diffraction-limited measurements. As before, we estimate our ideal photon count as

$$\widetilde{N}_{ideal,nd}[\widetilde{k_x}, \widetilde{k_y}] = \eta_{avg}^{-1}[\widetilde{k_x}, \widetilde{k_y}] N_{detected,nd}[\widetilde{k_x}, \widetilde{k_y}] \tag{S35}$$

where $\eta_{avg}[\widetilde{k_x}, \widetilde{k_y}]$ is an average efficiency expected across the measurement and accounts for a material efficiency (lowered by material absorption) and a coupling efficiency (which we average across all angles which contribute to the measurement)

$$\eta_{avg}[\widetilde{k_x}, \widetilde{k_y}] = \eta_{material}[\widetilde{k_x}, \widetilde{k_y}] \cdot AVG_{k_i \in \widetilde{k_i} \pm \Delta k_d}(\eta_{coupling}[k_x, k_y]) \tag{S36}$$

This efficiency $\eta_{avg}$ is our best guess as to the efficiency of a measurement. However, as explained, if the efficiency of the coupling varies significantly over the range of angles covered by a single measurement, it may not be exact. In contrast, we assume that $\eta_{coupling}[k_x, k_y]$ can be treated as constant over our diffraction-limited bin size. Then, our actual ideal photon count can be described as

$$N_{ideal,nd}[\widetilde{k_x}, \widetilde{k_y}] = \sum_{k_i \in \widetilde{k_i} \pm \Delta k_d} \eta^{-1}[k_x, k_y] N_{detected}[k_x, k_y]$$

Thus, we define a per-measurement normalization error which we assume is independent of shot noise

$$E[\widetilde{k_x}, \widetilde{k_y}] = N_{ideal,nd}[\widetilde{k_x}, \widetilde{k_y}] - \widetilde{N}_{ideal,nd}[\widetilde{k_x}, \widetilde{k_y}] \tag{S37}$$

where the $N_{ideal,nd}$ corresponds to the true ideal collection count, without the normalization error. Then, we modify Eqn. S26 to include this error our SNR

$$SNR = \frac{\sum_{\widetilde{k_x}, \widetilde{k_y}} \widetilde{N}_{ideal,nd}[\widetilde{k_x}, \widetilde{k_y}]^2}{\sum_{\widetilde{k_x}, \widetilde{k_y}} \eta_{avg}^{-1}[\widetilde{k_x}, \widetilde{k_y}] N_{ideal,nd}[\widetilde{k_x}, \widetilde{k_y}] + E^2[\widetilde{k_x}, \widetilde{k_y}]} \tag{S38}$$

We use this equation to generate the SNR points in Fig. 5b of the manuscript. We find that for small bin sizes, the shot noise contribution to the error dominates the system—in this regime, the total acquisition time is important. However, if the bins are made too wide, the normalization error dominates the SNR of the system and acquisition time ceases to matter.

In Fig. 2a of the manuscript, we consider only the effects of shot noise on SNR, because we simulate image acquisition in the diffraction-limited regime.

## 12. Normalization and the array factor calculation

As discussed in the manuscript, it is important in image retrieval to normalize the detected photon by the measurement efficiency. In our simulation, this normalization is performed by fixing the following condition to be true

$$\int_{\frac{\vec{k}_{in}}{|k_{in}|} \leq 1} |G(\vec{0}, \vec{k}_{in})|^2 |A(\vec{k}_{in})|^2 d\vec{k}_{in} = 1 \tag{S39}$$

This section justifies this normalization.

Assume that all light incident on the active metasurface couples to the scatterers and moreover assume that there are no material losses in the resonator. This is the condition which gives us the ideally retrieved photon count which we need for image reconstruction.

We consider first the case where a single wave of unit power is normally incident on the metasurface, such that $I_{in}(\vec{k}_{in}) = \delta(\vec{k}_{in})$. From Eqn. S1, we determine that

$$I_{detected}(\vec{k}_{out}) = |G(\vec{k}_{out}, \vec{0})|^2 |A(-\vec{k}_{out})|^2 \tag{S40}$$

Since we assume full coupling and no material losses, it follows that the total scatterer photon flux is equal to the incident photon flux. Thus,

$$1 = \int_{\frac{\vec{k}_{out}}{|k_{out}|} \leq 1} I_{detected}(\vec{k}_{out}) d\vec{k}_{out} = \int_{\frac{\vec{k}_{out}}{|k_{out}|} \leq 1} |G(\vec{k}_{out}, \vec{0})|^2 |A(-\vec{k}_{out})|^2 d\vec{k}_{out} \tag{S41}$$

By reciprocity, we know that $|G(\vec{k}_{out}, \vec{0})|^2 = |G(\vec{0}, \vec{k}_{out})|^2$. Additionally, we note that the array factor $|A(\vec{k})|^2 = R|A(-\vec{k})|^2$ where $R$ denotes a reflection $k_x \to -k_x, k_y \to -k_y$. Then, since our domain of integration is invariant to the reflection, we can write

$$1 = \int_{\frac{\vec{k}_{out}}{|k_{out}|} \leq 1} |G(\vec{0}, \vec{k}_{out})|^2 |A(\vec{k}_{out})|^2 d\vec{k}_{out} = \int_{\frac{\vec{k}_{in}}{|k_{in}|} \leq 1} |G(\vec{0}, \vec{k}_{in})|^2 |A(\vec{k}_{in})|^2 d\vec{k}_{in} \tag{S42}$$

Thus, we can use this normalization to find our ideal photon count, $N_{ideal}$.

## 13. Image reconstruction algorithms

We will describe the reconstruction algorithms here in terms of ideal photon count, $N_{ideal}$. The procedure by which the ideal photon count is found from the detected photon count $N_{detected}$ is presented in the manuscript.

In all imaging simulations, we simulate the far field coupling and far field scene at a higher resolution than the diffraction-limited resolution of the active metasurface. This allows us to treat our measurement results as continuous integrals presented in the manuscript.

**Point-by-point imaging.** Each measurement is assumed to come from a distinct wavevector in the far field. Then, each wavevector corresponds to a distinct point in space. We generate our image $I[k_x, k_y]$ as $I[k_x, k_y] = N_{ideal}[k_x, k_y]$ (up to normalization of the full image).

**Hadamard imaging.** The Hadamard basis is defined as a basis in k-space consisting of a $2^N \times 2^N$ element grid (for integer $N$), where each square can take on a value of either $+1$ or $-1$ in each basis element and all elements are orthogonal. Thus, retrieving the inner product between each basis element and the scene requires two measurements—one for the positive component and one for the negative component. Alternatively, one could take only the positive measurements and calculate the negative measurement from the first basis element (where each grid element is $+1$), halving the total number of required measurements. We opt for the first option in Fig. 3f of the manuscript.

The inner product of the scene with the first basis element is obtained in a single measurement, as that element is positive. For all other elements, we determine the weight attributed to a basis element $i$ from the positive and negative measurements $N_{ideal}^{p,i}, N_{ideal}^{n,i}$ as:

$$w_i = N_{ideal}^{p,i} - N_{ideal}^{n,i} \tag{S43}$$

Note that in time-critical applications, measurements should be taken in order of increasing spatial frequency[14].

**Edge-detection/kernel imaging.** We perform edge detection by taking a convolution between an edge-detection kernel and the scene. We understand this as taking the inner product between the far field and an optimized kernel, centering the kernel at each point of the scene. The inner product gives us the strength of an edge at a given point.

As with Hadamard imaging, we take two measurements to resolve the inner product at each point (one for the positive component of the kernel and another for the negative component). In the most general case where the positive and negative parts of the kernel have different weights, we must rescale $N_{ideal}^{p,i}$, $N_{ideal}^{n,i}$ before taking their difference. This is because the physical ideal measurement we take will give the same total weight to positive and negative coupling.

Assume that in our mathematically ideal kernel $K$, the ratio of the positive and negative coupling is

$$R_{coupling} = \frac{I_{pos}}{I_{neg}} = \frac{\max \int (K[k_x, k_y], 0) d\vec{k}}{\max \int (-K[k_x, k_y], 0) d\vec{k}} \tag{S44}$$

We then express the strength of an edge centered at $k_x, k_y$ as

$$w_{edge}[k_x, k_y] = N_{ideal}^{p,i}[k_x, k_y] I_{pos} - N_{ideal}^{n,i}[k_x, k_y] I_{neg} \tag{S45}$$

## 14. Gerchberg-Saxton optimization and enforcing metasurface properties

The metasurface properties are enforced by expressing the scatterers' amplitudes/phases given by the GS as points in the complex plane, and then casting each scatterer's response to the nearest physically achievable point in the complex plane, as defined by the L2 norm. The casting of arbitrary responses to their nearest achievable points generally depends on the choice of normalization for the scatterer amplitudes (the GS algorithm provides us with the relative but not absolute amplitudes). We chose to normalize the amplitudes of the GS algorithm such that the greatest amplitude was 0.37. This provided us with best convergence. An example of a converged set of points (blue) and their nearest achievable values (red) is provided in Fig. S5.

In the case of the Hadamard basis optimization, we found that the GS algorithm with enforced metasurface properties led to a large, undesirable coupling peak at $(k_x, k_y) = (0, 0)$. This coupling peak was reduced by adding a random perturbation to the designed voltages. The results shown in Fig. 3e and Fig. 3f of the manuscript were generated with a voltage perturbation uniformly distributed between -0.9 and 0.9V.

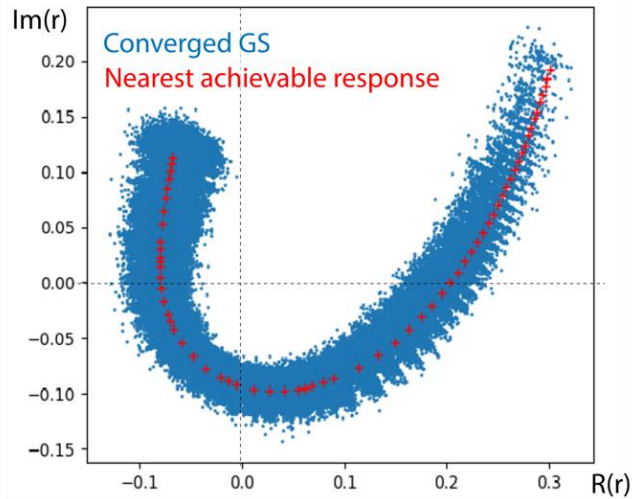

**Figure S5.** Complex response of $511^2$ individual scatterers as designed by the GS algorithm (blue) and achievable complex response (red) which was used in simulating metasurface performance.